\documentclass[sigconf]{acmart}
\PassOptionsToPackage{usenames,dvipsnames,table}{xcolor}

\usepackage{graphicx}

\usepackage{float}

\usepackage{pgfplots}
\pgfplotsset{compat=1.16}

\usepackage{amsmath,amsfonts}
\usepackage{algorithmic}
\usepackage{textcomp}
\usepackage{booktabs}
\usepackage[a-1b]{pdfx}
\usepackage{subcaption}
\usepackage{xcolor}
\usepackage{colortbl}
\usepackage{bbding}
\usepackage{braket}
\usepackage{epsfig,latexsym,graphics}
\usepackage[font={small}]{caption}
\usepackage{setspace}
\usepackage{enumitem}

\usepackage{cmd}
\usepackage[spacesep,definevectors]{easyvector} 
\usepackage{tabularx}
\usepackage{makecell}
\usepackage{multirow}
\usepackage{bbm} 
\usepackage{fontawesome5}
\usepackage[normalem]{ulem} 

\SetAlgoInsideSkip{0pt}

\copyrightyear{2026}
\acmYear{2026}
\setcopyright{cc}
\setcctype{by}
\acmConference[KDD 2026] {Proceedings of the 32nd ACM SIGKDD Conference on Knowledge Discovery and Data Mining V.2}{August 9--13, 2026}{Jeju Island, Republic of Korea.}
\acmBooktitle{Proceedings of the 32nd ACM SIGKDD Conference on Knowledge Discovery and Data Mining V.2 (KDD 2026), August 9--13, 2026, Jeju Island, Republic of Korea}
\acmISBN{979-8-4007-2259-2/2026/08}
\acmDOI{10.1145/3770855.3817860}

\begin{document}

\title{E2E: Efficient Filtered AKNN Search via Adaptive Termination}

\author{Wenxuan Xia}
\authornote{Both authors contributed equally to this research.}
\affiliation{%
  \institution{The Hong Kong University of Science and Technology (Guangzhou)}
  \city{Guangzhou}
  \state{Guangdong}
  \country{China}
}
\email{wxia248@connect.hkust-gz.edu.cn}

\author{Mingyu Yang}
\authornotemark[1]
\authornote{Corresponding author.}
\affiliation{%
  \institution{The Hong Kong University of Science and Technology (Guangzhou)}
  \city{Guangzhou}
  \state{Guangdong}
  \country{China}
}
\email{myang250@connect.hkust-gz.edu.cn}

\author{Wentao Li}
\affiliation{%
\institution{University of Leicester}
    \city{Leicester}
    \state{England}
  \country{United Kingdom}
}
\email{wl226@leicester.ac.uk}

\author{Wei Wang}
\authornotemark[2]
\affiliation{%
  \institution{The Hong Kong University of Science and Technology (Guangzhou)}
  \city{Guangzhou}
  \state{Guangdong}
  \country{China}
}
\affiliation{%
\institution{The Hong Kong University of Science and Technology}
  \city{Hong Kong}
  \country{Hong Kong}
}
\email{weiwcs@ust.hk}


\begin{abstract}
Approximate $k$-Nearest Neighbor (AKNN) search is widely used in vector databases.
When vectors carry additional attributes (e.g., labels or numerical values), \emph{filtered} AKNN search retrieves the nearest vectors to a query vector under attribute constraints.
Most existing methods use a fixed termination condition, searching the entire index while respecting attribute filters.
However, this leads to substantial redundant computations, since different queries require different amounts of search effort, and thus misses early termination opportunities for easy queries.
This paper proposes a lightweight model to estimate the search cost of filtered AKNN queries and enable \emph{adaptive termination}:
For easy queries, the search stops early to reduce latency, while for hard queries, it continues longer to preserve accuracy.
The key challenge is accurate cost prediction under attribute filters.
To address this, we show that information collected during an early probing phase (e.g., attribute distributions and intermediate distance statistics) can effectively predict the overall search cost.
Experiments on six real-world datasets demonstrate $1.1\times$--$3.7\times$ speedup over state-of-the-art baselines at $95\%$ recall, while maintaining search accuracy.
\end{abstract}




\maketitle

\vspace{-1em}
\section{Introduction}\label{sec:intro}
$k$-Nearest Neighbor (KNN) search aims to find the $k$ vectors closest to a query vector in a database.
It has wide applications in retrieval-augmented generation (RAG) for large language models (LLMs)~\cite{NN-datamining-1967-TIT,LLM-RAG-NIPS-2020}, data mining~\cite{Data-Mining-Concepts-Jiawei-Han}, and multi-modal retrieval \cite{MQA-VLDB-2024-Wang}.
However, exact KNN search is extremely costly in high-dimensional spaces, especially for large datasets.
Thus, approximate KNN (AKNN) search is proposed, which sacrifices some accuracy to significantly improve efficiency~\cite{Acorn-SIGMOD-2024}.

\begin{figure}[!t]
    \centering
    \includegraphics[width=\linewidth]{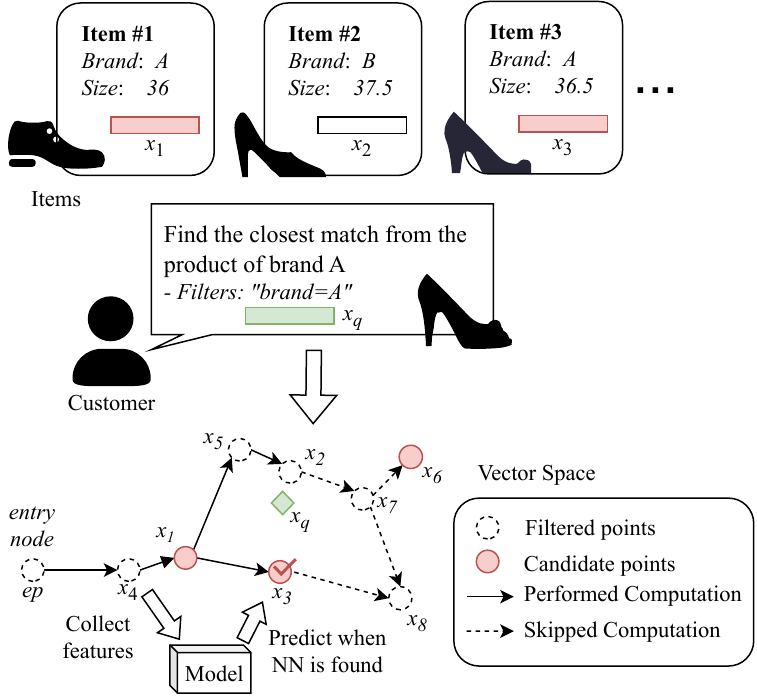}\vspace{-1em}
    \caption{An example of filtered AKNN search ($k=1$) in an e-commerce scenario, where vectors are associated with categorical label attributes.
A user submits a reference photo $x_q$ with the filter \texttt{Brand=A}.
As shown in the search process (bottom), some geometrically close neighbors (e.g., $x_2$ and $x_5$) are filtered out due to label mismatch.
While traditional methods may exhaustively traverse the graph to ensure recall, our framework, \EE, exploits signals from an early probing phase (e.g., visits to $x_1$, $x_4$, and $x_5$) to enable adaptive termination, stopping early once $x_3$ is found to improve efficiency.}
    \label{fig:intro-example}
\end{figure}

In recent years, many applications, especially in industry (e.g., recommender systems and LLMs), associate database vectors with attributes.
These attributes can be discrete (e.g., item categories) or continuous (e.g., numerical values such as prices)~\cite{Acorn-SIGMOD-2024,Vbase-OSDI-2023,AnalyticDB-VLDB-2020,VSAG-arxiv-2025}.
In such settings, plain AKNN search is insufficient when users impose constraints on attributes~\cite{NSG-VLDB-2019-deng-cai}.
As a result, \emph{filtered} AKNN search has been proposed, which incorporates attribute constraints alongside the query vector to retrieve the $k$ nearest neighbors.
It enables users to exclude unwanted vectors and perform AKNN search under attribute-based filters~\cite{ReconfigurableInvertedIndex-MM-2018-Matusi,MANSW-WILEY-2019,PASE-SIGMOD-2020-yang,ConstrainedApproximateSimilarity-Arxiv-2022-zhao,AnalyticDB-VLDB-2020,Milvus-SIGMOD-2021,NHQ-NIPS-2022-mengzhao-wang, HQANN-CIKM-2022, HQI-SIGMOD-2023,Vbase-OSDI-2023,Filtered-diskann-WWW-2023, ARKGraph-VLDB-2023-Zuo,Acorn-SIGMOD-2024, UNIFY-VLDB-2024-liang, SeRF-SIGMOD-2024,SuperPostfiltering-ICML-2024-Joshua,DIGRA-SIGMOD-2025-sibo-menxu-cuhk, Wow-Range-SIGMOD-2025, RWalks-SIGMOD-2025-Ait, UNG-SIGMOD-2025, ESG-arxiv-2025-Mingyu,Hi-PNG-KDD-2025-Ming-Yang, iRangeGraph-SIGMOD-2025,  DynamicRangefilteringApproximate-VLDB-2025-Peng, Beyond-Vector-Search-Jiadong-Xie-SIGMOD-2025,ELI-VLDB-2026-Mingyu,SIEVE-VLDB-2026, FilterAKNNBenchmark-SIGMOD-2025}.
Fig.~\ref{fig:intro-example} shows an example of filtered AKNN search, where only vectors satisfying the filter (e.g., $x_1, x_3, x_6$) are candidates for the results.

\stitle{Existing Solutions.}
To support filtered AKNN search, existing methods typically rely on graph-based indexes~\cite{HNSW-PAMI-2020,SSG-PAMI-2022-deng-cai,tMRNG:journals/pacmmod/PengCCYX23,HVS-VLDB-2021-kejing-lu,VSAG-arxiv-2025,DEG-SIGMOD-2025,SymphonyQG-SIGMOD-2025,AnalyticDB-VLDB-2020,CPG-SIGMOD-2026-Shangqi-Lu} due to their state-of-the-art performance.
These indexes were originally designed for plain AKNN search: they model vectors as nodes in a navigable graph, where each node connects to nearby neighbors.
Starting from an entry point, the search iteratively moves toward nodes closer to the query.
To support filtered AKNN search, existing methods mainly adopt two strategies, namely $\PRE$ and $\POST$, to enforce attribute constraints during traversal.
(1) The $\PRE$ strategy restricts the search space by filtering out nodes (and incident edges) that violate the attribute constraint during traversal.
Yet, this may disconnect and over-sparsify the graph, preventing the search from finding enough valid candidates to return $k$ neighbors.
(2) The $\POST$ strategy~\cite{hnswlib,Diskann-NIPS-2019,HNSW-PAMI-2020} traverses the \textit{full} graph but only inserts candidate vectors satisfying the attribute-based filters into the result set.
Given its superior performance, it has been widely adopted in industry.

\begin{figure}[!t]
    \centering
    \includegraphics[width=0.6\linewidth]{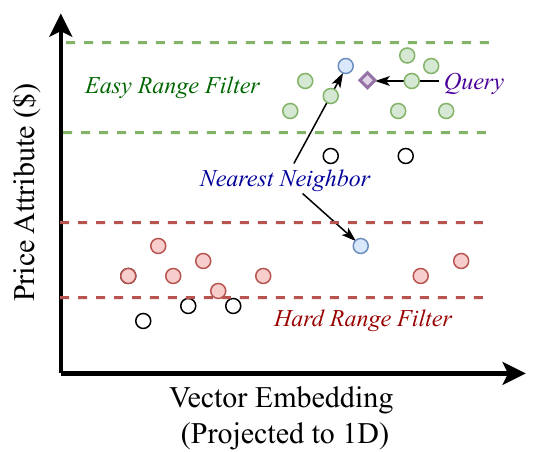}\vspace{-0.5em}
    \vgap\caption{Examples of easy and hard queries.
A filtered AKNN search query is \emph{easy} when filter-valid vectors lie near the query vector, enabling early termination once the nearest valid neighbor is found.
In contrast, a search is \emph{hard} when valid vectors are far away, requiring much more graph traversal.
As a result, search costs vary significantly across filtered AKNN search queries.}
    \label{fig:intro-motivation}
\end{figure}

\vspace{-0.3em}
\stitle{Limitations.}
Even when adapted for filtered AKNN search, both $\PRE$ and $\POST$ still rely on a graph-based index.
They start from an entry node and navigate the graph until the $k$ nearest neighbors are returned, typically using a fixed termination condition that is applied uniformly to all filtered queries.
However, the search cost varies significantly across different filters.
As illustrated in Fig.~\ref{fig:intro-motivation}, some \emph{easy} queries quickly encounter high-quality candidates that satisfy the attribute-based filters, requiring only limited exploration; in contrast, \emph{hard} queries may need to traverse a much larger portion of the graph before enough valid neighbors are found.
To guarantee high recall under such heterogeneous workloads, existing systems typically search the entire graph index~\cite{DARTH-SIGMOD-2026,HNSW-PAMI-2020}.
However, this strategy is inherently inefficient: it wastes substantial computation on easy queries, which are more common than hard ones in practice~\cite{LAET-SIGMOD-2020,DARTH-SIGMOD-2026}.

\stitle{Motivations.}
To reduce redundant exploration in graph-based AKNN search, prior work trains a learning-based model~\cite{LAET-SIGMOD-2020,DARTH-SIGMOD-2026} to predict query cost, where \emph{easy} queries require little cost while \emph{hard} queries demand much more.
With this model, existing methods search longer for hard queries to preserve accuracy, while stopping early for easy queries to improve efficiency.
The key novelty of these methods lies in designing features that accurately capture query cost.
For example, LAET~\cite{LAET-SIGMOD-2020} and DARTH~\cite{DARTH-SIGMOD-2026} monitor runtime signals (e.g., the distance distribution of visited vectors during graph traversal) to estimate query cost.
Yet, these techniques are designed for \emph{plain} AKNN search.
They typically assume that \textbf{true neighbors are typically discovered early during the search~\cite{DARTH-SIGMOD-2026}}, which is also required for graph-based indexes to work properly.
Under this assumption, distance-based signals are informative: if visited candidate vectors quickly become close to the query, the search cost is likely small.

In contrast, they cannot be directly applied to \emph{filtered} AKNN search.
The reason is that they ignore the attributes associated with vectors.
Even if many nearby vectors are visited (yielding small distances), they may be invalid due to filter mismatch.
As shown in Fig.~\ref{fig:intro-motivation}, for hard filtered AKNN search queries, valid results satisfying the attribute-based filters may lie far away.
Consequently, small geometric distances no longer imply low cost, since the search may still need to traverse a large portion of the graph to find enough valid neighbors.
As a result, existing cost prediction methods for plain AKNN can fail in our setting.

\stitle{Our Solution.}
To address the limitations of distance-only features for cost prediction in filtered AKNN search, we propose \textbf{\EE} (\underline{E}arly probe-to-\underline{E}arly termination).
The key idea is to incorporate the \emph{attribute distribution} of vectors via a low-cost early probing phase, beyond runtime distance signals.
Specifically, we introduce two filter-aware features:
(1) the \textit{observed valid ratio}, i.e., the fraction of \textit{visited} vectors satisfying the filter; and
(2) the \textit{prospective valid ratio}, i.e., the fraction of \textit{queued} vectors satisfying the filter.

The former captures \emph{local} selectivity in the explored neighborhood, while the latter reflects a more \emph{global} signal since further expansion is driven by the queue.
Both features are collected during the probing phase without additional overhead: the search already traverses the graph index, and we only record extra statistics.
We then combine them with existing distance-based features and train a lightweight tree-based model (LightGBM)~\cite{LightGBM-NIPS-2017} to predict per-query search cost.
To evaluate their effectiveness, we conduct a feature-importance analysis over all features used for cost prediction, including both the distance-based and our proposed filter-aware features.
As shown in Fig.~\ref{fig:misalign}, the results confirm that filter-aware features are especially critical when attribute distributions are skewed and misaligned with distance-based signals.
This confirms the need of incorporating features tailored to filtered AKNN search.

\stitle{Contributions.}
We summarize our main contributions as follows:

\sstitle{Problem Analysis of Existing Methods ($\S$~\ref{sec:problem-analysis}).}
We formally analyze the limitations of existing query cost prediction methods (e.g., LAET~\cite{LAET-SIGMOD-2020} and DARTH~\cite{DARTH-SIGMOD-2026}).
We show that relying solely on distance-based runtime features is insufficient for accurately estimating the cost of filtered AKNN search queries.

\sstitle{The $\EE$ Cost Estimation Framework ($\S$~\ref{sec:ee-frame}).}
To address this limitation, we propose \EE, to the best of our knowledge the first cost estimation framework dedicated to \emph{filtered} AKNN search. \EE incorporates attribute-distribution signals in addition to distance statistics: it actively samples the query neighborhood to extract filter-aware features, and feeds them into a lightweight prediction model to obtain more accurate cost estimates.

\sstitle{Application to Adaptive Termination ($\S$~\ref{sec:Apply}).}
We demonstrate how \EE can be integrated into existing graph-based indexes to enable adaptive termination for filtered AKNN search.
Specifically, \EE distinguishes easy queries from hard ones and adjusts the termination condition accordingly: easy queries stop early to improve efficiency, while hard queries search longer to preserve recall, guided by the predicted cost.

\sstitle{Extensive Evaluation ($\S$~\ref{sec:Exp}).}
We evaluate \EE on multiple large-scale real-world datasets with both categorical and numerical attributes.
Results show that \EE achieves up to a \textbf{3$\times$ reduction in search latency} over strong baselines at the same recall, and consistently outperforms state-of-the-art adaptive methods by correctly identifying hard, anti-correlated queries.
For reproducibility, our code is available at \url{https://github.com/gegeji/E2E}.

\section{Preliminary}\label{sec:preliminaries}
In this section, we first formalize the filtered AKNN search problem, and then review graph-based AKNN indexes and existing approaches for query cost estimation.

\subsection{Problem Statement}
We consider an \emph{attributed vector database}, where each vector is associated with either discrete or continuous filter attributes.

\begin{definition}[Attributed Vector Database]
An item in an attributed vector database $S$ is defined as $o_i=(x_i, A_i)$, where $x_i \in \mathbb{R}^d$ is a $d$-dimensional vector and $A_i$ denotes its filter attribute.
We consider two primary attribute types:
\begin{itemize}[leftmargin=4mm]
    \item \textbf{Discrete labels:} $A_i \subseteq \mathcal{L}$, where $\mathcal{L}$ is the global label alphabet.
    \item \textbf{Continuous values:} $A_i \in \mathbb{R}$, i.e., a continuous scalar attribute.
\end{itemize}
\end{definition}

For simplicity, we assume that all items in $S$ share the same attribute type (either discrete or continuous), and we do not consider hybrid settings.
When the context is clear, we refer to items in $S$ simply as \emph{vectors}.
Given an attributed vector database $S$, a filtered query is denoted as $q=(x_q,f_q)$, where $x_q$ is the query vector and $f_q$ is the \textit{attribute constraint} (or \textit{filter} for short).
Compared to plain AKNN search, filtered queries impose additional attribute constraints to exclude mismatched vectors.
The constraint $f_q$ can be a label set $f_q=L_q$ or a numerical range $f_q=[l,r]$.
To determine whether an item satisfies a query, we define a \textit{filter predicate}.

\begin{definition}[Filter Predicate]
Let $o_i=(x_i,A_i)\in S$.
A filter predicate $\mathcal{P}(A_i,f_q)\in\{\text{True},\text{False}\}$ evaluates whether the attribute $A_i$ of item $o_i$ satisfies the query filter $f_q$.
We consider the following predicate types:
\begin{itemize}[leftmargin=4mm]
    \item \textbf{Label Containment:}
    $\mathcal{P}_{\text{contain}}(A_i,f_q)=\text{True}\iff f_q \subseteq A_i$,
    where $f_q$ is a label set (i.e., $f_q=L_q$).
    
    \item \textbf{Label Equality:}
    $\mathcal{P}_{\text{equal}}(A_i,f_q)=\text{True}\iff A_i=f_q$,
    where $f_q$ is a label set (i.e., $f_q=L_q$).
    
    \item \textbf{Numerical Range:}
    $\mathcal{P}_{\text{range}}(A_i,f_q)=\text{True}\iff A_i\in f_q$,
    where $f_q$ is a numerical range (i.e., $f_q=[l,r]$).
\end{itemize}
\end{definition}

Note that many other predicate types are possible; we use the above three as representative examples.
Given a database $S$, a query $q$, and a predicate logic $\mathcal{P}$, the filtered subset is defined as:
\[
S(q) = \{ o_i \in S \mid \mathcal{P}(A_i,f_q) = \text{True} \} .
\]

\begin{definition}[Filtered $k$-Nearest Neighbor (KNN) Search]
Given an attributed vector database $S$ and a query $q=(x_q,f_q)$, let
$S(q)$ denotes the set of data items satisfying the filter.
The \emph{filtered KNN search} problem is to return a set $R \subseteq S(q)$ with $|R|=k$ such that
for any $o_i \in R$ and $o_j \in S(q)\setminus R$,
$\delta(x_i,x_q) \le \delta(x_j,x_q)$,
where $\delta(\cdot,\cdot)$ is the distance function.
\end{definition}

Since an exact KNN search is prohibitively expensive at scale, we relax the requirement of returning the exact neighbors and instead allow approximation.
This leads to the \emph{filtered approximate $k$-nearest neighbor} (AKNN) query $q=(x_q, f_q)$, which returns the approximate $k$NNs of $x_q$ subject to the attribute constraint $f_q$.

\stitle{Selectivity.}
For a filtered query $q=(x_q, f_q)$, a natural indicator of query difficulty is the size of the filtered subset
$S(q)$, which contains all data items in the database $S$ that satisfy the filter condition.
To capture this notion more generally, we define \emph{selectivity} as the fraction of items in a given candidate set that match the filter.

\begin{definition}[Selectivity]
Given a query $q=(x_q, f_q)$ and a target set of data items $C \subseteq S$, the selectivity $\sigma(C, f_q)$ is defined as the proportion of items in $C$ that satisfy the filter:
\[
\sigma(C, f_q) = \frac{|C \cap S(q)|}{|C|}.
\]
\end{definition}

Depending on the scope of $C$, we further distinguish two types of selectivity:

\begin{itemize}[leftmargin=4mm]
    \item \textbf{Global Selectivity ($\sigma_{\text{global}}$).}
    When $C=S$ (the entire database), $\sigma(S, f_q)$ measures the probability that a randomly chosen item satisfies the filter.
    This value is often available from database statistics.
    
    \item \textbf{Local Selectivity ($\sigma_{\text{local}}$).}
    When $C$ is a subset of $S$, e.g., the proximity neighborhood of $x_q$ (such as the top-$m$ nearest neighbors under vector distance),
    $\sigma(C, f_q)$ measures selectivity within this local region.
\end{itemize}

\subsection{AKNN Search Methods}
Exact $k$-Nearest Neighbor (KNN) search often suffers from the \textit{curse of dimensionality}, causing traditional indexing methods~\cite{beckmannRtreeEfficientRobust,ann-benchmakrs} to degrade significantly in high-dimensional spaces. Consequently, most recent work has focused on \emph{Approximate} KNN search, which improves efficiency with only a small loss in accuracy. Among AKNN methods, graph-based indexes are the most popular due to their state-of-the-art query performance\cite{HNSW-PAMI-2020, NSG-VLDB-2019-deng-cai, Diskann-NIPS-2019, tMRNG:journals/pacmmod/PengCCYX23,Graph-Revisited-TODS-2025-Jiadong-Xie}.

\vspace{-0.3em}

\stitle{Graph Indexes}. 
Graph-based indexes model the database $S$ as a navigable proximity graph, where each vector is a node connected to its nearby neighbors constructed by dedicated algorithms~\cite{SSG-PAMI-2022-deng-cai,HVS-VLDB-2021-kejing-lu,HNSW-PAMI-2020,tMRNG:journals/pacmmod/PengCCYX23,CPG-SIGMOD-2026-Shangqi-Lu,Acorn-SIGMOD-2024}. 
Most graph indexes perform search using a greedy traversal strategy for top-$1$ nearest neighbor retrieval~\cite{NSG-VLDB-2019-deng-cai,Diskann-NIPS-2019,HNSW-PAMI-2020,Worse-case-NIPS-2023}. 
Specifically, the search starts from a designated entry point $ep$ and iteratively explores neighbors that are closer to the query until convergence.
To retrieve the top-$k$ neighbors, graph indexes typically employ a beam search procedure.
We denote the beam width as $m$ (i.e., $efSearch$ in HNSW~\cite{ESG-arxiv-2025-Mingyu}), which controls the size of the candidate queue.
A larger $m$ generally improves recall by expanding more nodes, but it also increases latency.

\vspace{-0.3em}
\stitle{Filtered Queries.}
When the items in a vector database $S$ are associated with attributes, \emph{filtered} AKNN search aims to retrieve the $k$ nearest neighbors of a query vector that also satisfy an attribute-based filter.
To support this functionality, existing approaches typically build on top of graph-based indexes and incorporate filtering during traversal.
In general, two representative strategies have been widely adopted~\cite{Acorn-SIGMOD-2024,iRangeGraph-SIGMOD-2025,UNG-SIGMOD-2025,DIGRA-SIGMOD-2025-sibo-menxu-cuhk,RangePQ-SIGMOD-2025-fangyuan-sibo,Wow-Range-SIGMOD-2025}: $\PRE$ and $\POST$.
Both strategies keep the index structure unchanged and enforce the filter at query time.
Specifically, given a filtered query $q=(x_q,f_q)$:
(1) $\PRE$ expands only nodes that satisfy $f_q$, i.e., it prunes filter-mismatch nodes (and their outgoing edges) during traversal;
(2) $\POST$ performs traversal without considering $f_q$, but discards filter-mismatch nodes when inserting candidates into the result set.
For top-$k$ queries, $\POST$ continues collecting nearest candidates until $k$ valid results are obtained.
Since $\PRE$ may disconnect the graph and lead to insufficient valid candidates, most systems adopt $\POST$ in practice~\cite{hnswlib,Diskann-NIPS-2019}, at the cost of computing distances for a large number of filtered-out points.

\subsection{Cost Estimation for AKNN Search}
To tune graph-based AKNN search, one can adjust search parameters (e.g., the beam size).
However, most existing systems still rely on a \emph{fixed} termination condition, i.e., they traverse the index until the search budget is exhausted, leaving little room for early stopping.
To enable early termination, lightweight learning-based methods such as LAET~\cite{LAET-SIGMOD-2020} and DARTH~\cite{DARTH-SIGMOD-2026} estimate the query cost required to reach a target recall.
These methods primarily exploit runtime signals, especially the distance distribution of visited candidates.
Their key intuition is that true neighbors are typically discovered early during the search~\cite{DARTH-SIGMOD-2026}; Otherwise, a large portion of the graph must be traversed, which makes graph-based indexing inefficient.
Therefore, by collecting signals from the early stage, the system can predict the remaining cost and stop once the estimated budget is reached.
However, we will show that cost estimation methods designed for \emph{plain} AKNN search are not suitable for analyzing the cost of \emph{filtered} AKNN queries in $\S$~\ref{sec:problem-analysis}.

\vspace{-0.5em}

\section{Problem Analysis}\label{sec:problem-analysis}
In the previous introduction, we argued that existing learning-based cost estimation methods do not work well for \emph{filtered} AKNN search.
In this section, we first provide a detailed analysis of the cost composition of filtered AKNN search, and then summarize the key challenges for accurate cost estimation.

\vspace{-1em}

\subsection{Cost Analysis of Filtered AKNN Search}


\stitle{Plain AKNN Search.}
We first analyze the search complexity of a graph-based AKNN index (e.g., HNSW) under \emph{plain} search, i.e., without considering filters.
The search consists of two phases.

\sstitle{Greedy Routing.}
This phase quickly locates the first approximate nearest neighbor of the query vector $x_q$ in the graph.
Starting from an entry node, the algorithm greedily moves to a neighbor closer to $x_q$ until no closer node exists.
The routing cost is typically $O(\log N)$, where $N$ is the database cardinality.
Advanced indexes such as HNSW~\cite{HNSW-PAMI-2020} and HVS~\cite{HVS-VLDB-2021-kejing-lu} further reduce this cost via hierarchical layers or sparse navigational graphs, making routing overhead negligible compared to the subsequent expansion phase.

\sstitle{Local Expansion.}
After obtaining an approximate top-1 neighbor, the search switches to a local exploration phase, commonly implemented via beam search.
Specifically, it iteratively expands the frontier by visiting neighbors of the currently explored nodes, while maintaining a candidate \textbf{queue} of size $m$ (i.e., the beam width).
For a standard top-$k$ search without filtering, the overall complexity is often expressed as $O(\log N + m \cdot \bar{d})$, where $\bar{d}$ is the average node degree.
Since the maximum degree in graph-based indexes is usually bounded by a constant (e.g., $M_{\max}$ in $\HNSW$), the cost of expanding a single node is effectively $O(1)$.
Therefore, the total complexity is dominated by the number of expansions needed to fill the beam, and is commonly simplified to $O(\log N + m)$~\cite{Acorn-SIGMOD-2024,ESG-arxiv-2025-Mingyu}.

\stitle{Filtered AKNN Search.} 
Given a filtered AKNN query $q=(x_q,f_q)$, where $x_q$ is the query vector and $f_q$ is the filter,
the $\POST$ strategy follows the standard two-stage search procedure of graph indexes.
The key difference is in the second stage: after locating the top-1 neighbor of $q$ (identical to plain AKNN search without considering attributes), the algorithm incrementally expands neighbors to collect $m$ candidates that satisfy $f_q$.
A major drawback of $\POST$ is that it still performs many distance computations for vectors that do not satisfy $f_q$, leading to high overhead.
Assuming independence between data distribution and filtering predicates, the expected cost to retrieve $m$ valid neighbors is $O(m/\sigma_{\text{global}})$~\cite{Acorn-SIGMOD-2024,ESG-arxiv-2025-Mingyu},
where $\sigma_{\text{global}}$ denotes the global selectivity.
Therefore, the overall complexity of $\POST$ is
\begin{equation}\label{eq:cost-model-learned}
    O(\log N + m/\sigma_{\text{global}}).
\end{equation}
\vspace{-2em}

\begin{figure}[t]
    \centering
    \includegraphics[width=0.8\linewidth]{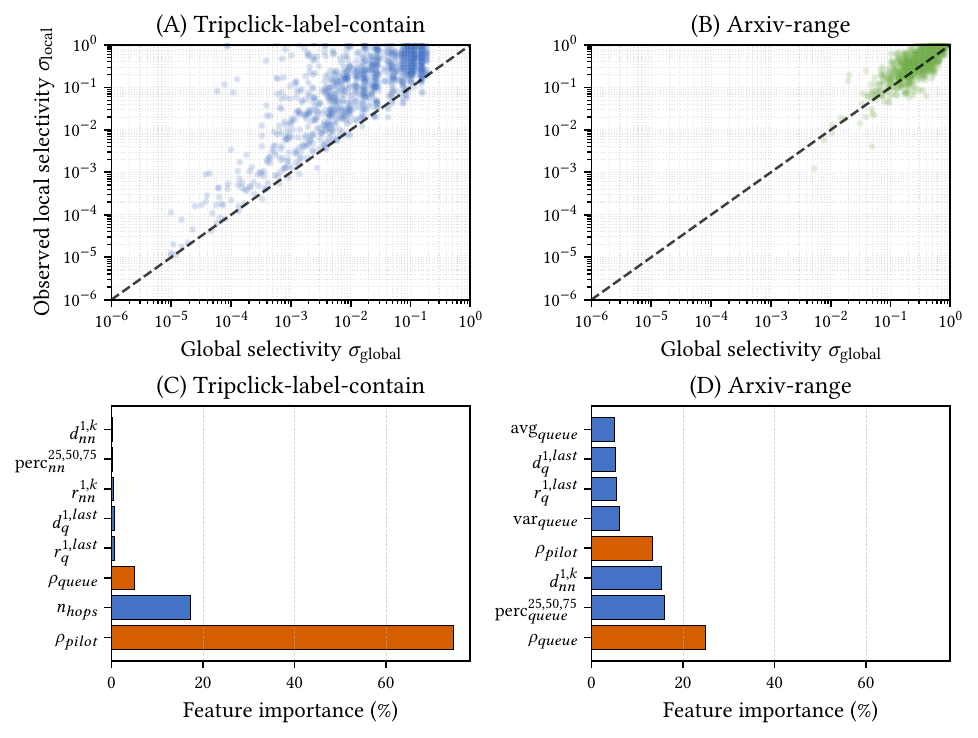}\vspace{-1em}
    \vgap\caption{
    Relationship between local ($\sigma_{\text{local}}$) and global ($\sigma_{\text{global}}$) selectivity.
(A) and (B) report results on Tripclick-label-contain and Arxiv-range: (A) shows little correlation, while (B) exhibits a strong linear trend.
(C) and (D) show feature importance for cost prediction, where blue denotes distance-based features and yellow denotes ours.
(C) shows that when local and global selectivity are misaligned, distance-only features perform poorly, motivating our filter-aware features; (D) shows that when they are aligned, both distance-only and our features are effective.
Overall, these results confirm that misalignment occurs in real datasets and motivates incorporating new features to handle such cases.
   }\vgap
    \label{fig:misalign}\vspace{0.5em}
\end{figure}

\vspace{-1em}

\subsection{Challenges}
At first glance, the cost of filtered AKNN search seems to be accurately estimated by Eq.~(1), since all parameters—such as the database size $N$, beam width $m$, and global selectivity $\sigma_{\text{global}}$—are known in advance.
However, this estimate relies on the assumption that the vector distribution is independent of the attribute distribution.
In real-world attributed vector databases, this assumption often fails.
Instead, vectors and their associated attributes can exhibit strong positive or negative correlations.
To verify this, we conduct a statistical analysis over diverse queries on two real-world datasets, Tripclick-label-contain and Arxiv-range.
For each query, we compute the global selectivity $\sigma_{\text{global}}$ (fraction of vectors in the dataset satisfying the filter) and the local selectivity $\sigma_{\text{local}}$ (fraction of valid vectors among the top-$m$ nearest neighbors).
As shown in Fig.~\ref{fig:misalign} (A), $\sigma_{\text{local}}$ can deviate greatly from $\sigma_{\text{global}}$, breaking the original hypothesis.
Thus, the neighbor-expansion cost in the $\POST$ stage depends not on $\sigma_{\text{global}}$, but on the query-specific $\sigma_{\text{local}}$.

\stitle{Challenges.} 
What are the consequences of the mismatch between global selectivity and local selectivity?

\sstitle{Challenge $\# 1$.}
Recall that existing cost estimators for \emph{plain} AKNN search (e.g., LAET~\cite{LAET-SIGMOD-2020} and DARTH~\cite{DARTH-SIGMOD-2026}) assume that early-stage distance signals can predict the final query cost, and thus use these early distance statistics to estimate the total cost.
In filtered AKNN search, however, cost is no longer purely distance-driven: the search must also find enough \emph{valid} candidates satisfying the filter.
When local selectivity aligns with global selectivity, early observations remain representative and distance-only features may still correlate with cost.
In real attributed vector databases, attribute--vector correlations can cause local selectivity to deviate from global selectivity, biasing the early explored neighborhood.
Thus, distance features become unreliable: a query may appear easy yet incur high cost due to scarce valid vectors, or appear hard but terminate quickly after reaching a high-selectivity region.
Therefore, \textit{plain AKNN cost estimators that ignore local--global selectivity mismatch are unsuitable for predicting the cost of filtered AKNN queries}.

\sstitle{Challenge $\# 2$.}
While selectivity estimation~\cite{SE-probalistic-SIGMOD-2001,SE-range-predicae-VLDB-2019,SelNet-SIGMOD-2021-Wang,GL-SIGMOD-2021} has been studied for decades, existing methods primarily estimate \emph{global} attribute density, which is insufficient for filtered AKNN search.
Moreover, filtered AKNN requires sub-millisecond response times, making state-of-the-art learned estimators impractical: their complex deep models often incur inference overheads that exceed the search time itself.
This motivates new methods for \emph{local} selectivity estimation, tailored to filtered AKNN cost prediction.

\begin{figure*}[!t]
    \centering
    \includegraphics[width=0.8\textwidth]{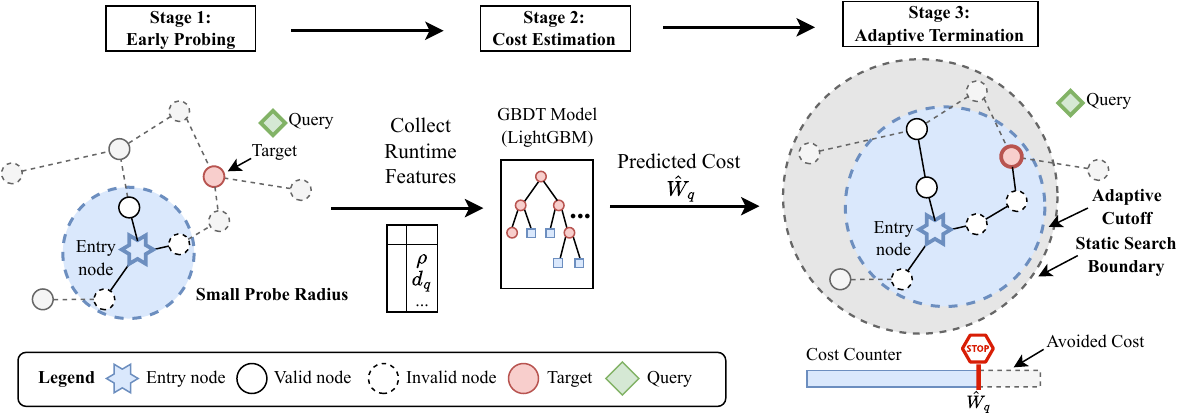}
    \vgap \vspace{-1em}
    \caption{The overall architecture of our proposed $\EE$ framework.}\vgap
    \label{fig:main_framework}\vgap
\end{figure*}

\section{Our Proposed Cost Estimation Framework}\label{sec:Model-Train}
As discussed in $\S$~\ref{sec:problem-analysis}, filtered AKNN search often suffers from a mismatch between global and local selectivity, which prevents existing cost estimators from working reliably.
To address this challenge, we design a new cost prediction framework $\EE$ tailored for filtered AKNN search, explicitly accounting for the local--global selectivity misalignment.
We first present an overview in Section~\ref{sec:ee-frame}, introduce our features in Section~\ref{sec:features}, describe model training in Section~\ref{sec:Training}, and integrate $\EE$ into graph-based search in Section~\ref{sec:Apply}.

\vspace{-1em}
\subsection{Framework Overview}\label{sec:ee-frame}
Our proposed framework, \EE, is tailored for filtered AKNN search, and its overall workflow is illustrated in Fig.~\ref{fig:main_framework}.
Recall that graph-based search typically consists of two stages: greedy routing and local expansion.
Greedy routing incurs only a small overhead, whereas local expansion dominates the query cost.
Motivated by this observation, \EE leverages the information collected during the first stage (greedy routing), referred to as the \emph{early probing module}, to extract runtime features and predict the query cost via the \emph{cost estimation module}.
Based on the predicted cost, \EE further enables early stopping in the second stage (local expansion) through an \emph{adaptive termination module}.
Next, we briefly introduce these three modules and then present their details in the following subsections.

\stitle{Early Probing.}
In the first stage of graph-based search (i.e., greedy routing), the query starts from an entry node and explores nearby nodes within a limited probe radius (illustrated by the blue dashed circle of Fig.~\ref{fig:main_framework}).
During this process, we collect two types of runtime signals: (i) \emph{selectivity-aware} features derived from the visited vectors (defined in Section~\ref{sec:features}), and (ii) the distance-based features used in prior work~\cite{LAET-SIGMOD-2020,DARTH-SIGMOD-2026}.
Note that greedy routing is an indispensable component of graph-based indexes, as it is required to locate the first approximate nearest neighbor.
Our contribution is to extract lightweight features from this mandatory stage, which incurs negligible overhead.
Concretely, we record the first $f$ distance computations of the \emph{actual} query execution during early probing (typically tens to a few hundreds), where $f$ is calibrated via a binary search procedure from full-search traces to meet recall targets and in practice the predictor is broadly insensitive to $f$, e.g., on Arxiv-contain,  $f\in[42,132]$ yield comparable $R^2$ ($0.35$--$0.40$) and NDC.
Moreover, if the estimator predicts that the target cost/recall requirement has already been satisfied, we directly return (or reuse) the candidates collected in this phase.

\stitle{Cost Estimation.}
Based on the features extracted in the early probing module, we train a lightweight model (i.e., a GBDT-based cost estimator implemented with LightGBM) to predict the query cost $\hat{W}_q$.
We adopt a tree-based estimator to ensure both low inference overhead and high accuracy (see Section~\ref{sec:Training}).

\vspace{-0.3em}

\stitle{Adaptive Termination.}
Recall that in the first stage of graph search, we perform early probing to extract runtime features and train a model to predict the query cost.
In the second stage (local expansion), the search must continue exploring the graph to collect enough nearest neighbors that satisfy the filter, which is often the dominant cost.
Fortunately, our cost predictor enables \emph{adaptive early termination}: we dynamically monitor the cost incurred so far and stop the expansion once it reaches the predicted budget $\hat{W}_{q}$.
This allows us to prune unnecessary node expansions and directly return the current results.

\vspace{-0.3em}

\stitle{Remark.}
Given a query $q$, we quantify its cost, denoted as $W_q$, by the number of distance computations (NDCs).
We use NDCs as our primary metric because it is hardware-agnostic and correlates well with search latency, while abstracting away machine and implementation details.

\vspace{-1.5em}

\subsection{Feature Extraction}\label{sec:features}
The next question is how to extract effective features for filtered AKNN search.
As analyzed in Section~\ref{sec:problem-analysis}, when local and global selectivity are aligned, the original distance-based features remain informative (Fig.~\ref{fig:misalign}).
However, when they are misaligned, distance-only features become insufficient.

\stitle{Distance-Based Features.}
We first review the distance-based features used in prior work, such as LAET~\cite{LAET-SIGMOD-2020} and DARTH~\cite{DARTH-SIGMOD-2026}.
In graph-based search, we maintain (i) a candidate \textbf{queue} $\mathcal{Q}$ of nodes to explore and (ii) a \textbf{result set} $\mathcal{R}$ of current AKNN answers.
Prior studies extract features from both structures, along with global search statistics collected during traversal.
Specifically, we extract three types of distance-based features in Table~\ref{tbl:features}.

\noindent
$\bullet$
\textbf{(1) Global features}, including the distance from the entry node to the query, and the total number of search pops.

\noindent
$\bullet$
\textbf{(2) Queue features}, including the distances of nodes in the candidate queue to the query, the distance ratio of queued nodes, and summary statistics of queue distances (mean, variance, and the 25/50/75 percentiles).

\noindent
$\bullet$
\textbf{(3) Result-set features}, including the distances of nodes in the current result set to the query, the distance ratio of result nodes, and summary statistics of result-set distances (mean, variance, and the 25/50/75 percentiles).

\begin{table}[!t]
\centering
\caption{Summary of Input Features for the Predictor.}\vgap
\label{tbl:features}
\footnotesize
\begin{tabularx}{\linewidth}{l|l|X}
\toprule
\textbf{Category} & \textbf{Notation} & \textbf{Description} \\
\midrule
\multirow[t]{2}{*}{Global} 
 & $d_{start} {}^\dagger$ & Distance from the entry node to the query. \\
 & $n_{hops} {}^\ddagger$ & Total number of search hops. \\
\hline
\multirow[t]{5}{*}{Queue} 
 & $d_{queue}^{1, last} {}^\ddagger$ & Distances from the head and tail of the queue to the query.  \\
 & $r_{queue}^{1, last} {}^\dagger$ & Distance ratios in the queue, normalized by $d_{\textit{start}}$.  \\
 & $\text{avg}_{queue}, \text{var}_{queue} {}^\ddagger$ & Mean and variance of distances in the queue.  \\
 & $\text{perc}_{queue}^{25, 50, 75} {}^\ddagger$ &  25th, 50th, and 75th percentiles of distances in the queue.  \\
\hline
\multirow[t]{5}{*}{Result Set} 
 & $d_{nn}^{1, last} {}^* {}^\ddagger$ & Distances from the closest and farthest valid candidates in $\mathcal{R}$ to the query.  \\
 & $r_{nn}^{1, last} {}^\dagger$ & Distance ratios in $\mathcal{R}$, normalized by $d_{\textit{start}}$.  \\
 & $\text{avg}_{nn}, \text{var}_{nn} {}^\ddagger$ & Mean and variance of distances in $\mathcal{R}$.  \\
 & $\text{perc}_{nn}^{25, 50, 75} {}^\ddagger$ & 25th, 50th, and 75th percentiles of distances in $\mathcal{R}$.  \\
\bottomrule
\multicolumn{3}{p{\linewidth}}{\footnotesize $^\dagger$: Adapted from LAET \cite{LAET-SIGMOD-2020}, $^\ddagger$: Adapted from DARTH \cite{DARTH-SIGMOD-2026}.}
\end{tabularx}
\end{table}

\vspace{-0.5em}

\stitle{Filter-Aware Features.}
We next introduce a set of \emph{filter-aware} features tailored for filtered AKNN search.
The key intuition is that distance-based features remain effective when global and local selectivity are aligned (Fig.~\ref{fig:misalign}); otherwise, the cost is also governed by the attribute distribution encountered during traversal.
Accordingly, we treat selectivity (i.e., the fraction of visited vectors satisfying the filter) as a first-class signal.
Recall that Early Probing provides rich visitation statistics.
Based on these observations, we design local-selectivity features from two complementary perspectives:
(i) the \emph{history}, i.e., the vectors already visited; and
(ii) the \emph{future}, i.e., the vectors currently in the queue for subsequent expansions.
Specifically, we define the following two features.

\noindent
$\bullet$
\textbf{Observed valid ratio $\rho_{pilot}$.}
We define $\rho_{pilot}$ as the fraction of filter-matching vertices visited during the early probing stage:
$\rho_{pilot} = \frac{N_{\text{valid\_visited}}}{N_{\text{total\_visited}}}$,
where $N_{\text{valid\_visited}}$ is the number of valid nodes among the visited nodes in early probing, and $N_{\text{total\_visited}}$ is the total number of visited nodes in this stage.
We use $\rho_{pilot}$ as a proxy for the local selectivity $\sigma_{\text{local}}$ for the query.
Unlike global selectivity, $\rho_{pilot}$ indicates if the query lies in a region that is locally dense with valid nodes, or in a sparse region where valid matches are scarce.

\noindent
$\bullet$
\textbf{Prospective valid ratio $\rho_{queue}$.}
This metric measures the fraction of filter-matching vertices in the priority queue:
$\rho_{queue} = \frac{N_{\text{valid\_in\_queue}}}{N_{\text{queue}}}$,
where $N_{\text{valid\_in\_queue}}$ is the number of valid nodes in the queue, and $N_{\text{queue}}$ is the queue size.
Intuitively, $\rho_{queue}$ reflects the expected difficulty of the immediate future expansions.
A low $\rho_{queue}$ implies that, although the queued candidates are close to $x_q$, most of them are invalid (i.e., strong local--global misalignment), suggesting that a larger search budget is required to traverse these invalid candidates and expand the search scope sufficiently to retrieve the target $k$ valid neighbors.

\vspace{-1em}
\subsection{Model Training}\label{sec:Training}\vspace{-0.5em}

\stitle{Prediction Model.}
After extracting the features, the next step is to choose a model for cost prediction.
We adopt a Gradient Boosting Decision Tree (GBDT) model~\cite{LightGBM-NIPS-2017}, which constructs an ensemble by sequentially adding decision trees, where each new tree is trained to fit the residual errors of the current ensemble.
We implement our estimator using LightGBM~\cite{LightGBM-NIPS-2017}, a highly optimized GBDT framework.
We choose LightGBM over alternatives such as XGBoost~\cite{XGboost-KDD-2016} primarily for its superior inference efficiency in single-output regression: our trained model requires only $0.025$\,ms per query on average in a single-threaded setting, occupies roughly $624$\,KB on 1M dataset, and incurs no per-vertex or per-edge storage.
This low-latency property is crucial for our adaptive framework, ensuring that repeatedly evaluating the termination condition introduces negligible overhead compared to the overall query execution time.

\stitle{Training Strategy.}
We adopt a supervised learning paradigm to train the cost estimator.
To construct the training set, we replay query logs on a hold-out dataset.
For each query $q$, we first run a standard greedy beam search with a sufficiently large queue to extract the runtime feature vector $z_q$ (including both distance-based and filter-aware features).
We then continue the search until the \emph{exact} top-$k$ neighbors are retrieved (i.e., 100\% recall), and record the total number of distance computations at convergence as the regression target $W_q$.
This $k$-NN-grounded label differs from LAET~\cite{LAET-SIGMOD-2020}, whose target is the cost to reach a \emph{single} nearest neighbor; the distinction matters under attribute-vector misalignment, where the $k$-th valid neighbor can lie much farther than the closest one, and LAET would ignore the cost from reaching the closest neighbor to retrieving all valid $k$ nearest neighbors.
This procedure yields a labeled dataset $\mathcal{D}=\{(z_q, W_q)\}_{q \in S_{\text{train}}}$, which maps the intermediate search state to the total cost required to reach the target recall.

Unlike step-wise predictors (e.g., DARTH~\cite{DARTH-SIGMOD-2026}) that extract multiple training samples from a single trajectory, our end-to-end formulation provides one scalar label per query.
As a result, this formulation may require a larger query pool to match the sample volume of step-wise prediction approaches \cite{DARTH-SIGMOD-2026}. In practice, however, the model plateaus at $\sim$50k training queries (Fig.~\ref{fig:convergence} in Appendix A.4), and model training itself takes less than $5$ seconds on all datasets.
Moreover, ground-truth generation is a one-time offline procedure taking under $30$ minutes per $1$M training vector workload and is highly parallelizable, making it feasible even at scale.

\begin{algorithm}[!t]
	\caption{$\POST$ Search with Early Probing and Adaptive Termination}
	\label{algo:search-algorithm-integration}
	\begin{footnotesize}
	\KwIn{Query $q=(x_q, f_q)$, filter predicate $\mathcal{P}(\cdot, f_q)$, number of target nearest neighbors $k$, entry point $ep$, early probing budget $f$, tolerance factor $\alpha$, graph $G$, cost estimator $\mathcal{M}$}
    \KwOut{The result set $\mathcal{R}$}
    $\mathcal{R} \leftarrow$ empty max heap with size $k$ as result set\;
    $\mathcal{Q} \leftarrow$ min heap candidate queue\;
    $\mathcal{V} \leftarrow$ empty visited set\; 
    $\mathcal{Q}$.push($ep$); $cnt \leftarrow 0$\tcp*[r]{$cnt$ counts NDC}
    \tcc{Early probing phase}
    \While{$cnt < f$ and $\mathcal{Q}$ is not empty}{
        $u \gets \mathcal{Q}$.pop()\;
        \For{$v \in N(u)$ }{\tcp{$v$ is one of $u$'s out-neighbors $N(u)$}
            \If{$v \notin \mathcal{V}$}{
                $\mathcal{Q}$.push($v$); $\mathcal{V}$.add($v$); $cnt \leftarrow cnt +1$\;
                \If{$\mathcal{P}(A_v, f_q)$ is $True$} {
                \tcp{$A_v$ is attribute value of $v$} $\mathcal{R}$.update($v$)\;}
            }
        }
    }
    Extract features $z_q$ (e.g., $\rho_{queue}, \rho_{pilot}$) from $\mathcal{Q}, \mathcal{R}$ and search history\;
    $\hat{W}_q \leftarrow \mathcal{M}(z_q) $\;
    \tcc{Adaptive Termination phase}
    \While{$cnt < \alpha \cdot \hat{W}_q$ and $\mathcal{Q}$ not empty}{
       $u \gets \mathcal{Q}$.pop() \;
        \For{$v \in N(u)$ }{
            \If{$v \notin \mathcal{V}$}{
                $\mathcal{Q}$.push($v$); $\mathcal{V}$.add($v$); $cnt \leftarrow cnt +1$\;
                \If{$\mathcal{P}(A_v, f_q)$ is $True$}{$\mathcal{R}$.update($v$)\;}
            }
        }
    }
    
    \textbf{return} top-$k$ nearest neighbors in $\mathcal{R}$
\end{footnotesize}
\end{algorithm}

\stitle{Log-Based Loss Function.}
Directly regressing the raw cost $W_q$ is challenging due to the \textit{heavy-tailed} cost distribution in filtered AKNN search.
The number of distance computations (i.e., $W_q$) can vary by orders of magnitude (e.g., from $10^2$ to $10^6$), largely depending on filter selectivity.
Training on such highly skewed targets causes the loss to be dominated by hard queries with large $W_q$, which often harms generalization on easy queries.
To address this issue, we follow prior work~\cite{LAET-SIGMOD-2020} and apply a logarithmic transformation to the target.
Specifically, the model $\mathcal{M}$ is trained to predict $\log(W_q)$ by minimizing the Mean Squared Error (MSE) in log-space:
\begin{equation}
\mathcal{L} = \sum_{q \in S_{\text{train}}} \bigl(\mathcal{M}(z_q) - \log(W_q)\bigr)^2.
\end{equation}
This objective is equivalent to minimizing the \textit{squared log-ratio error},
$(\log(W_q) - \log(\hat{W}_q))^2 = \bigl(\log \frac{W_q}{\hat{W}_q}\bigr)^2$.
Unlike standard MSE, which emphasizes absolute errors, this formulation implicitly optimizes the \textit{relative error}.
As a result, the model remains sensitive to prediction deviations for both easy and hard queries, yielding robust cost estimates across the full difficulty spectrum.

\subsection{Integrated into Graph-based Search}\label{sec:Apply}
We now integrate our cost prediction model into the graph-based index. 
Algorithm~\ref{algo:search-algorithm-integration} summarizes how we embed the proposed adaptive early termination mechanism into the graph search procedure. We implement on top of HNSW \cite{HNSW-PAMI-2020, hnswlib} as it is one of the widely adopted graph-based indexes. We note that specialized filter-aware index designs \cite{Acorn-SIGMOD-2024, Filtered-diskann-WWW-2023, NHQ-NIPS-2022-mengzhao-wang, UNG-SIGMOD-2025} are orthogonal to our cost prediction model: they redesign the index structure to handle filters, whereas $\EE$ decides \emph{when to stop} on top of any such index. We demonstrate composition E2E with ACORN in Appendix A.3.

\stitle{Algorithm.}
The search is initialized from the entry point $ep$.
After initialization (Lines 1--4), we trigger an early probing phase.
Specifically, we run a beam search where each visited node is also evaluated against the query filter $f_q$ (Lines 5--11).
This phase is constrained by a small fixed budget $f$, i.e., a pre-defined number of distance computations.
Once the budget is exhausted, we pause the traversal and extract runtime features (Line 12), capturing both (i) the local behavior of $f_q$ in the explored neighborhood and (ii) the current search progress.
These features are then fed into our predictor (Line 13) to estimate the total search cost $\hat{W}_q$.

The algorithm then switches to the adaptive termination phase (Lines 14--20).
To flexibly trade efficiency for recall, we introduce a tolerance factor $\alpha$ to calibrate the stopping threshold.
The traversal resumes from the preserved state (i.e., maintaining $\mathcal{V}$ and $\mathcal{R}$) and continues until the cumulative number of distance computations reaches $\alpha \cdot \hat{W}_q$.
This adaptive budget prevents the over-computation common in static configurations, where easy queries may unnecessarily consume excessive distance evaluations (e.g., $cnt \mathrel{>\!\!>} k/\sigma_{\text{local}}$).
We calibrate $\alpha$ for a target recall by a binary search over a small set of sampled validation queries (under $1$ second), and recall increases monotonically with $\alpha$ (the calibrated values typically lie in $[1.1, 20]$ depending on the dataset and target recall, see Figures.~\ref{fig:recall-query-latency-main}--\ref{fig:recall-avg-dist-comp-main}).

Finally, the algorithm terminates and returns the top-$k$ approximate nearest neighbors that satisfy $f_q$ (Line 21).

\input{figures/qps_recall_plot/query_latency_vs_recall}

\input{figures/qps_recall_plot/avg_dist_comp_vs_recall}

\begin{figure}
    \centering
    \begin{small}
    
    \begin{tikzpicture}
    \begin{customlegend}[legend columns=4,
    legend entries={$\EE$, Naive HNSW, $\EE$ w/o Filter},
    legend style={at={(0.5,1.15)},anchor=north,draw=none,font=\scriptsize,column sep=0.3cm}]
    \addlegendimage{line width=0.2mm,color=violate,mark=o,mark size=0.8mm}
    \addlegendimage{line width=0.2mm,color=amaranth,mark=triangle,mark size=0.8mm}
    \addlegendimage{line width=0.2mm,color=navy,mark=square,mark size=0.8mm}
    \end{customlegend}
    \end{tikzpicture}
    \\[-\lineskip]
    \subfloat[Tripclick-label-equality]{\vspace{-2mm}
    \begin{tikzpicture}[scale=1]
    \begin{axis}[
        height=\columnwidth/2.75,
        width=0.230\textwidth, 
        xlabel=Query Latency (ms), xlabel style={yshift=+3pt},
        ylabel=Recall@10, ylabel style={yshift=-4pt}, 
        label style={font=\scriptsize},
        tick label style={font=\scriptsize},
        ymajorgrids=true,
        xmajorgrids=true,
        grid style=dashed,
        title={Avg. Select 1\%},
        title style={font=\scriptsize},
        title style={yshift=-2.5mm},
    ]   
    \addplot[line width=0.2mm,color=violate,mark=o,mark size=0.6mm]
    plot coordinates {
        (11.5110, 94.3213)
        (13.5710, 94.6648)
        (15.5210, 95.1869)
        (17.5150, 95.5910)
        (19.4870, 95.9047)
        (21.4880, 96.2077)
        (23.5290, 96.3188)
        (25.6110, 96.4198)
        (27.6730, 96.5006)
        (31.9260, 96.5410)
        (29.7780, 96.6218)
        (34.0040, 96.6623)
        (36.1330, 96.7835)
        (40.4880, 96.8755)
        (38.1881, 96.9249)
        (45.6621, 97.1684)
        (51.4941, 97.3199)
        (57.0910, 97.5321)
        (69.2118, 97.5624)
        (63.4469, 97.6331)
        (75.3898, 97.8250)
        (108.4300, 97.8295)
        (102.1070, 97.8396)
        (82.1652, 97.8856)
        (95.3107, 97.9103)
        (121.7749, 97.9720)
        (88.4502, 97.9911)
        (114.6450, 98.1134)
        (128.0010, 98.1134)
    };
    \addplot[line width=0.2mm,color=amaranth,mark=triangle,mark size=0.6mm]
    plot coordinates {
       (58.7610, 93.6639)
        (84.7709, 95.8357)
        (100.9750, 96.6942)
        (115.6900, 97.1589)
        (125.2420, 97.4922)
        (136.7809, 97.7750)
        (167.5241, 97.9165)
        (172.9039, 97.9367)
        (147.1701, 97.9468)
        (159.0000, 98.1488)
    };
    \addplot[line width=0.15mm, color=navy, mark=square, mark size=0.5mm] 
    plot coordinates {
        (16.7620, 93.4190)
        (19.9170, 94.0958)
        (22.1720, 94.2877)
        (24.4260, 94.5402)
        (26.7070, 94.7119)
        (31.2880, 94.9535)
        (28.9890, 94.9804)
        (33.5910, 95.1881)
        (38.2610, 95.2941)
        (40.6001, 95.4052)
        (35.9340, 95.4860)
        (42.9430, 95.7655)
        (45.2689, 95.9069)
        (47.0250, 96.1089)
        (49.2451, 96.1190)
        (58.9918, 96.1796)
        (54.1621, 96.2706)
        (51.5719, 96.3312)
        (56.4879, 96.3312)
        (63.1900, 96.3817)
        (67.7989, 96.5343)
        (65.5630, 96.5545)
        (71.7350, 96.6061)
        (60.8409, 96.6140)
        (71.5861, 96.6353)
        (70.1218, 96.7161)
        (74.0872, 96.7374)
        (77.1933, 96.8890)
        (79.4691, 96.9496)
        (81.4989, 96.9900)
    };
    \end{axis}
    \end{tikzpicture}}\hspace{1mm}%
    \subfloat[Arxiv-label-equality]{\vspace{-2mm}
    \begin{tikzpicture}[scale=1]
    \begin{axis}[
        height=\columnwidth/2.75,
        width=0.230\textwidth, 
        xlabel=Query Latency (ms), xlabel style={yshift=+3pt},
        ylabel=Recall@10, ylabel style={yshift=-4pt}, 
        label style={font=\scriptsize},
        tick label style={font=\scriptsize},
        ymajorgrids=true,
        xmajorgrids=true,
        grid style=dashed,
        title={Avg. Select 10\%},
        title style={font=\scriptsize},
        title style={yshift=-2.5mm},
    ]   
    \addplot[line width=0.2mm,color=violate,mark=o,mark size=0.6mm]
    plot coordinates {
       (2.0010, 88.2132)
        (3.9810, 95.4924)
        (5.7120, 97.6447)
        (6.7200, 98.6294)
        (6.6270, 99.1472)
        (7.6170, 99.4315)
        (8.8150, 99.6447)
        (10.0390, 99.7462)
        (11.1750, 99.7462)
        (12.3990, 99.8173)
    };
    \addplot[line width=0.2mm,color=amaranth,mark=triangle,mark size=0.6mm]
    plot coordinates {
        (53.1220, 90.2944)
        (57.5109, 93.9086)
        (56.2601, 95.9797)
        (56.4771, 97.1777)
        (55.0819, 97.9492)
        (54.6780, 98.4569)
        (55.1341, 98.7005)
        (55.5651, 98.9137)
        (55.9550, 99.0457)
    };
    \addplot[line width=0.15mm, color=navy, mark=square, mark size=0.5mm] 
    plot coordinates {
        (6.9530, 96.4772)
        (7.7890, 96.9137)
        (6.6310, 97.1777)
        (7.9950, 97.3706)
        (7.4660, 97.4924)
        (7.2720, 97.6447)
        (7.5380, 97.7665)
        (7.8930, 97.8782)
        (7.8780, 98.0000)
        (8.2700, 98.0711)
        (8.6700, 98.1523)
        (9.0610, 98.2030)
        (9.4590, 98.2741)
        (9.8420, 98.3249)
        (10.2470, 98.4061)
        (10.6380, 98.4162)
        (11.0420, 98.4569)
        (11.4410, 98.5178)
        (11.8500, 98.5685)
        (12.2300, 98.6091)
    };
    \end{axis}
    \end{tikzpicture}}\hspace{1mm}%
\vspace{-0.4cm}
\caption{The performance under low selectivity.}
\label{fig:equality-recall-latency}
\end{small}
\end{figure}

\section{Experiments}\label{sec:Exp}
\subsection{Experimental Setup}
We evaluate our framework on four real-world datasets spanning diverse domains, including text retrieval, audio recommendation, and citation analysis.
Based on the filter attribute type, we group these datasets into two benchmarks: \textit{Label-Filtered} (categorical labels) and \textit{Range-Filtered} (discrete numeric values).
A statistical summary is provided in Table~\ref{tbl:dataset}.

\stitle{Filter Query Generation.}
A core challenge in learning-based cost estimation is ensuring that training and testing queries share a consistent feature distribution.
To evaluate the generalization of our estimator, we construct the base database ($S$), the training set ($S_{\text{train}}$), and the testing query set ($S_{\text{test}}$, 1{,}000 items) as three mutually disjoint subsets of the source corpus for each dataset. When the source corpus is much larger than the target dataset size (e.g., MSMARCO with 113M vectors), all three subsets are drawn as disjoint random samples directly from the source. When the source corpus is comparable to $S$ in size (e.g., TripClick with $\sim$1.05M vectors), we instead randomly reserve $S_{\text{train}}$ and $S_{\text{test}}$ first and treat the remainder as $S$. Per-dataset sizes are listed in Table~\ref{tbl:workloads} (Appendix~A.2).
This design keeps the query distribution aligned with the underlying data manifold.
Construction details are provided in Appendix~A.2.

\begin{table}[t]
\centering
\caption{Statistics of Datasets and Filter Attributes.}
\vspace{-0.2cm}
\label{tbl:dataset}
\small
\begin{tabular}{l|c|c|c|c}
\toprule
\textbf{Dataset} & \textbf{Dim} & \textbf{\# Base} & \textbf{Filter Type} & \textbf{Attributes} \\
\midrule
Tripclick & 768  & 1.0M & Label & Clinical Areas \\
Youtube   & 128  & 1.0M & Label & Audio Tags \\
Arxiv     & 4096 & 1.7M & Label / Range & Categories / Date \\
MSMARCO   & 1024 & 1.0M & Range & Synthetic Int \\
\bottomrule
\end{tabular}
\end{table}

\stitle{Metrics.}
We evaluate our method $\EE$ from two perspectives:
(1) \emph{End-to-End Search Performance.}
We measure the trade-off between search quality and efficiency.
Search quality is measured by Recall@$k$, defined as the fraction of the true top-$k$ ground-truth neighbors that are retrieved.
Search efficiency is measured by query latency (in ms) and the number of distance computations (NDCs).
(2) \emph{Estimator Accuracy.}
To evaluate the learned cost model, we report the coefficient of determination ($R^2$), Spearman's correlation $\rho$, and the root mean squared error (RMSE) on the test set.

All experiments are conducted on a server equipped with two Intel Xeon Platinum 8352V CPUs @ 2.10\,GHz (36 cores, 72 threads in total) and 512\,GB of main memory.
To accelerate preprocessing, we construct the index using 128 threads.
In contrast, all search-time evaluations are performed on a single thread.
To reduce noise, we report the average results over three independent runs.
We further exclude the raw high-dimensional query vector from the feature set to keep inference latency low.
Due to space limitations, the experiment on extending $\EE$ to other AKNN methods (Exp-4) is deferred to the Appendix (\S~A.4).

\subsection{Experimental Result}

\stitle{Exp-1: Overall Performance.} 
We compare $\EE$ against Naive HNSW and a variant of our model \emph{without} filter-aware features (i.e., using only the distance-based features adopted from state-of-the-art cost estimators LAET~\cite{LAET-SIGMOD-2020} and DARTH~\cite{DARTH-SIGMOD-2026}; see Table~\ref{tbl:features} for details).
We evaluate the trade-off between search quality (Recall@$k$) and efficiency (query latency), as shown in Fig.~\ref{fig:recall-query-latency-main}.
We further report the trade-off between search quality and efficiency measured by the number of distance computations, with results presented in Fig.~\ref{fig:recall-avg-dist-comp-main}.
Overall, $\EE$ consistently outperforms the baselines on all attributed datasets, achieving a better recall--cost trade-off.
In the high-recall regime, $\EE$ achieves up to a $2.6\times$ speedup on Tripclick (see Fig.~\ref{fig:recall-query-latency-main}), and reduces distance computations by up to $3\times$ on YoutubeAudio (see Fig.~\ref{fig:recall-avg-dist-comp-main}), compared to Naive HNSW.
These results confirm that the gains mainly come from our adaptive early termination, rather than hardware-dependent effects.
Moreover, $\EE$ remains robust under low-selectivity settings, making it practical for challenging filtered search workloads. The smallest gain corresponds to the attribute-vector aligned datasets (e.g., $1.04\times$ at recall=95\% for Arxiv-range over \EE w/o Filter of Fig.~\ref{fig:misalign}), when the filter is decorrelated from the vector proximity, e.g., uniformly random query ranges over native numerical attributes.

\stitle{Exp-2: Accuracy of Cost Prediction.}
Table~\ref{tbl:model_perf_k10} reports the accuracy of our cost estimator.
The results reveal a clear divergence between regression quality (measured by $R^2$, ranging from 0.39 to 0.55) and ranking quality (measured by Spearman's $\rho$, reaching 0.79).
The moderate $R^2$ is expected, as local--global selectivity misalignment introduces substantial variance in the absolute search cost.
In contrast, the high rank correlation (e.g., 0.79 on MSMARCO) is the key success factor: it shows that $\EE$ reliably captures the \emph{relative} hardness of queries.
Finally, the tolerance factor $\alpha$ (introduced in Alg.~\ref{algo:search-algorithm-integration}) mitigates scale bias in absolute predictions.
By calibrating $\hat{W}_q$ with $\alpha$, the system allocates computation based on relative difficulty---spending more on hard queries and less on easy ones---despite noise in the absolute cost.

\begin{table}[!t]
\centering
\caption{The accuracy of cost prediction ($K=10$).}\vgap
\label{tbl:model_perf_k10}
\small
\setlength{\tabcolsep}{4pt}
\begin{tabular}{l|c|ccc}
\toprule
\textbf{Dataset} & \textbf{Filter} & \textbf{Log-RMSE} & \textbf{$R^2$} & \textbf{Spearman $\rho$} \\
\midrule
\textbf{Tripclick} & Equality & 2.54 & 0.42 & \textbf{0.76} \\
\textbf{Tripclick} & Contain. & 2.68 & 0.45 & \textbf{0.74} \\
\textbf{Arxiv} & Equality & 1.49 & 0.55 & \textbf{0.73} \\
\textbf{Arxiv} & Contain. & 1.47 & 0.39 & 0.63 \\
\textbf{Youtube} & Contain. & 2.38 & 0.52 & 0.54 \\
\textbf{Arxiv} & Range & 1.43 & 0.43 & 0.65 \\
\textbf{MSMARCO} & Range & 1.41 & 0.54 & \textbf{0.79} \\
\bottomrule
\end{tabular}\vspace{0.5em}
\end{table}

\stitle{Exp 3: Robustness under Low Selectivity.}
We further evaluate $\EE$ under \textit{equality filters}, with results shown in Fig.~\ref{fig:equality-recall-latency}.
In this setting, the query exhibits extremely low local selectivity.
As shown in Fig.~\ref{fig:equality-recall-latency}, Naive HNSW incurs high latency (e.g., $>50$ms on Arxiv) due to near-exhaustive traversal, whereas $\EE$ identifies the low local validity signal ($\rho_{pilot}\approx 0$) and dynamically reduces the search budget.
Consequently, $\EE$ achieves an order-of-magnitude lower latency at high recall, demonstrating strong robustness under highly selective workloads.

\vspace{1em}

\section{Related Work}
\vspace{-0.5em}

\stitle{Learning for AKNN Search.}
Learning-based approaches have achieved remarkable success in AKNN search, supporting a wide range of tasks such as vector quantization~\cite{OPQ-PAMI-2014,Learned-PQ-2020-ICML-guoruiqi}, distance computation~\cite{BSA-DDC-ICDE-2024-Mingyu}, cost estimation~\cite{PMLSH-bolong-2020}, graph routing~\cite{learning-to-roate2019}, and query optimization~\cite{Vbase-OSDI-2023}.
However, for \emph{filtered} AKNN search, existing methods do not explicitly model the interaction between vector attributes and query predicates.
Our proposed $\EE$ bridges this gap through a low-cost early probing phase that captures filter-aware signals, thereby substantially improving cost estimation accuracy.

\vspace{-0.3em}

\stitle{Filtered AKNN Search.}
Filtered AKNN search enables flexible attribute-based filtering over unmatched vectors and has attracted increasing attention.
For example, VBase~\cite{Vbase-OSDI-2023}, Chase~\cite{CHASE-arxiv-Sean-Wang}, and AnalyticDB~\cite{AnalyticDB-VLDB-2020} support filtered queries efficiently via query optimization.
ACORN~\cite{Acorn-SIGMOD-2024} and Filter-DiskANN~\cite{Filtered-diskann-WWW-2023} improve graph-based indexes to better handle general filtered search.
NHQ~\cite{NHQ-NIPS-2022-mengzhao-wang} accelerates filtered search by rewriting the distance operator and incorporating attributes into the similarity computation.
More specialized methods, including Window-Filter~\cite{Window-Filter-ICML-2024}, SeRF~\cite{SeRF-SIGMOD-2024}, iRange~\cite{iRangeGraph-SIGMOD-2025}, ESG~\cite{ESG-arxiv-2025-Mingyu}, WoW~\cite{Wow-Range-SIGMOD-2025}, UNG~\cite{UNG-SIGMOD-2025}, ELI~\cite{ELI-VLDB-2026-Mingyu}, and SIEVE~\cite{SIEVE-VLDB-2026}, build multi-graph indexes for range and label-filtered search, but often at the cost of substantially higher index space and construction time.

\vspace{-0.8em}

\section{Conclusion}
This paper studies adaptive termination for filtered AKNN search.
We first analyze the limitations of existing learning-based cost estimators, motivating our new framework $\EE$ for more accurate estimation.
In particular, we show that the misalignment between distance signals and attribute distributions is the key reason why prior methods fail, and propose filter-aware features as a remedy.
We extract features from an early probing stage, train a lightweight cost predictor, and integrate it into graph-based search for adaptive termination.
Extensive experiments confirm the effectiveness of our estimator.
Our framework is general beyond HNSW and applies to other graph-based indexes, while extending to IVF-based indexes requires extra modification, which is non-trivial and remains our future work. Adapting $\EE$ to handle query-distribution shifts and data updates is another direction for future work.


\begin{acks}
We are grateful to the anonymous reviewers for their constructive comments. Wentao Li is supported by NSFC (Grant No.\ 62302417). This work is supported by Advanced Materials---National Science and Technology Major Project (Grant No.\ 2025ZD0620100), National Key R\&D Program of China (No.\ 2024YFA1012700), and Guangdong Provincial Key Lab of Integrated Communication, Sensing and Computation for Ubiquitous Internet of Things (No.\ 2023B1212010007).
\end{acks}

\bibliographystyle{ACM-Reference-Format}
\bibliography{reference}

\begin{figure*}[!t]
    \centering
   
    \begin{small}
    
    \begin{tikzpicture}
    \begin{customlegend}[legend columns=3,
    legend entries={\EE-ACORN, ACORN, \EE-ACORN w/o Filter},
    legend style={at={(0.5,1.15)},anchor=north,draw=none,font=\scriptsize,column sep=0.3cm}]
    \addlegendimage{line width=0.2mm,color=violate,mark=o,mark size=0.8mm}
    \addlegendimage{line width=0.2mm,color=amaranth,mark=triangle,mark size=0.8mm}
    \addlegendimage{line width=0.2mm,color=navy,mark=square,mark size=0.8mm}
    \end{customlegend}
    \end{tikzpicture}
    \\[-\lineskip]
    
	\subfloat[Arxiv-label-contain]{\vspace{-2mm}
	\makebox[0.19\textwidth][c]{%
	\begin{tikzpicture}[scale=1, trim axis left, trim axis right]
	\begin{axis}[
	    height=0.08\textwidth,
	    width=0.15\textwidth,
	    scale only axis,
	    xlabel=Query Latency (ms), xlabel style={yshift=+3pt},
	    ylabel=Recall@10, ylabel style={yshift=-4pt}, 
	    label style={font=\scriptsize},
	    tick label style={font=\scriptsize},
	    ymajorgrids=true,
	    xmajorgrids=true,
	    grid style=dashed,
	]
    \addplot[line width=0.2mm,color=violate,mark=o,mark size=0.6mm]
    plot coordinates {
        (11.3890, 99.4700)
        (12.8130, 99.6000)
        (14.2060, 99.7200)
        (15.5840, 99.7700)
        (16.9710, 99.8400)
        (18.3580, 99.9100)
        (19.7570, 99.9200)
        (21.1490, 99.9600)
        (22.5540, 99.9700)
        (23.9440, 99.9900)
    };
    \addplot[line width=0.2mm,color=amaranth,mark=triangle,mark size=0.6mm]
    plot coordinates {
        (9.5363, 99.4900)
        (10.9973, 99.6500)
        (12.3723, 99.7300)
        (13.6973, 99.7900)
        (14.9957, 99.8300)
        (16.2577, 99.8600)
        (17.4640, 99.9200)
        (18.6443, 99.9300)
        (20.8707, 99.9500)
        (21.9507, 99.9800)
    };
    \addplot[line width=0.15mm, color=navy, mark=square, mark size=0.5mm] 
    plot coordinates {
        (11.9160, 99.3500)
        (12.6410, 99.5200)
        (12.2160, 99.5900)
        (13.6520, 99.7500)
        (15.4000, 99.8600)
        (16.7250, 99.8800)
        (18.1750, 99.8900)
        (19.6680, 99.9400)
        (23.9820, 99.9500)
        (25.5430, 99.9600)
        (31.4140, 99.9800)
    };
	\end{axis}
	\end{tikzpicture}}}%
	\hspace{1mm}
    %
    %
	\subfloat[Tripclick-label-contain ]{\vspace{-2mm}
	\makebox[0.19\textwidth][c]{%
	\begin{tikzpicture}[scale=1, trim axis left, trim axis right]
	\begin{axis}[
	    height=0.08\textwidth,
	    width=0.15\textwidth,
	    scale only axis,
	    xlabel=Query Latency (ms), xlabel style={yshift=+3pt},
	    ylabel=Recall@10, ylabel style={yshift=-4pt}, 
	    label style={font=\scriptsize},
	    tick label style={font=\scriptsize},
	    ymajorgrids=true,
	    xmajorgrids=true,
	    grid style=dashed,
	]
    \addplot[line width=0.2mm,color=violate,mark=o,mark size=0.6mm]
    plot coordinates {
        (4.7560, 79.3500)
        (5.5940, 79.7400)
        (6.4360, 80.0100)
        (8.8380, 80.1300)
        (7.2410, 80.2600)
        (10.3830, 80.3200)
        (11.9030, 80.3600)
        (14.1020, 80.4200)
        (14.8230, 80.5300)
        (15.5310, 80.6800)
        (15.9040, 80.7100)
        (16.5170, 80.8400)
    };
    \addplot[line width=0.2mm,color=amaranth,mark=triangle,mark size=0.6mm]
    plot coordinates {
       (3.7070, 79.4600)
        (6.0843, 79.7800)
        (8.1887, 79.8600)
        (17.1710, 80.0700)
        (41.9541, 80.1600)
        (60.7378, 80.1800)
        (61.9030, 80.4700)
        (68.4261, 80.6400)
    };
    \addplot[line width=0.15mm, color=navy, mark=square, mark size=0.5mm] 
    plot coordinates {
        (4.7300, 79.4400)
        (5.5340, 80.1800)
        (8.0490, 80.3900)
        (9.6820, 80.4100)
        (11.1150, 80.4800)
        (13.3210, 80.5200)
        (14.7530, 80.5600)
        (15.4830, 80.5900)
    };
	\end{axis}
	\end{tikzpicture}}}%
	\hspace{1mm}%
	\subfloat[Arxiv-label-equality ]{\vspace{-2mm}
	\makebox[0.19\textwidth][c]{%
	\begin{tikzpicture}[scale=1, trim axis left, trim axis right]
	\begin{axis}[
	    height=0.08\textwidth,
	    width=0.15\textwidth,
	    scale only axis,
	    xlabel=Query Latency (ms), xlabel style={yshift=+3pt},
	    ylabel=Recall@10, ylabel style={yshift=-4pt}, 
	    label style={font=\scriptsize},
	    tick label style={font=\scriptsize},
	    ymajorgrids=true,
	    xmajorgrids=true,
	    grid style=dashed,
	]
    \addplot[line width=0.2mm,color=violate,mark=o,mark size=0.6mm]
    plot coordinates {
        (13.4720, 97.9500)
        (16.1280, 98.1200)
        (18.7670, 98.2200)
        (21.3630, 98.2900)
        (24.1560, 98.3400)
        (26.8110, 98.3500)
    };
    \addplot[line width=0.2mm,color=amaranth,mark=triangle,mark size=0.6mm]
    plot coordinates {
       (19.2547, 97.9700)
        (23.8593, 98.3200)
        (30.3400, 98.3400)
        (36.1296, 98.3800)
        (41.1460, 98.3800)
        (45.6377, 98.3900)
    };
    \addplot[line width=0.15mm, color=navy, mark=square, mark size=0.5mm] 
    plot coordinates {
        (13.9510, 97.8900)
        (16.9760, 98.1000)
        (20.2850, 98.2400)
        (23.4030, 98.3000)
        (26.5180, 98.3300)
        (32.6490, 98.3700)
    };
	\end{axis}
	\end{tikzpicture}}}%
	\hspace{1mm}%
	\subfloat[MSMARCO-range]{\vspace{-2mm}
	\makebox[0.19\textwidth][c]{%
	\begin{tikzpicture}[scale=1, trim axis left, trim axis right]
	\begin{axis}[
	    height=0.08\textwidth,
	    width=0.15\textwidth,
	    scale only axis,
	    xlabel=Query Latency (ms), xlabel style={yshift=+3pt},
	    ylabel=Recall@10, ylabel style={yshift=-4pt}, 
	    label style={font=\scriptsize},
	    tick label style={font=\scriptsize},
	    ymajorgrids=true,
	    xmajorgrids=true,
	    grid style=dashed,
	]
    \addplot[line width=0.2mm,color=violate,mark=o,mark size=0.6mm]
    plot coordinates {
        (4.8760, 95.8400)
        (5.7050, 96.8500)
        (6.6890, 97.3700)
        (7.6340, 97.7800)
        (8.2610, 98.1000)
        (8.9000, 98.3100)
        (9.9160, 98.5200)
        (10.6830, 98.6700)
        (11.6550, 98.8000)
        (12.6220, 98.9100)
        (13.3430, 99.0000)
        (14.3490, 99.0300)
        (15.2420, 99.1300)
        (16.0810, 99.2000)
        (17.2780, 99.2200)
    };
    \addplot[line width=0.2mm,color=amaranth,mark=triangle,mark size=0.6mm]
    plot coordinates {
        (4.3257, 94.8400)
        (5.8770, 96.1600)
        (7.4066, 96.8500)
        (8.5033, 97.3700)
        (9.6267, 97.6700)
        (10.7567, 97.9800)
        (11.7780, 98.1200)
    };
    \addplot[line width=0.15mm, color=navy, mark=square, mark size=0.5mm] 
    plot coordinates {
        (3.9920, 95.3500)
        (4.7860, 96.5600)
        (5.7320, 97.1400)
        (7.3720, 97.5800)
        (7.7390, 97.8700)
        (8.7610, 98.1400)
        (10.1380, 98.3400)
        (11.1760, 98.5100)
        (12.2200, 98.6300)
        (13.2450, 98.7500)
        (14.1530, 98.8400)
        (14.8470, 98.8800)
        (16.0840, 98.9600)
        (17.0120, 99.0600)
        (18.1260, 99.1100)
    };
	\end{axis}
	\end{tikzpicture}}}%
	\hspace{1mm}%

    \vspace{-0.4cm}
    \caption{Performance comparison between \EE{}-ACORN-$\gamma$ and ACORN-$\gamma$.
Top: Recall@10 vs. query latency (ms).
Bottom: Recall@10 vs. average distance computations (NDCs).}
    \label{fig:recall-latency-acorn-gamma}
    \end{small}
    
\end{figure*}

\balance

\section*{A Appendix}

\subsection*{A.1 Notations}
We summarize the commonly used notation in Table \ref{tbl:notation}.

\begin{table}[H]
  \small
  \caption{Commonly used notation used in this paper.} \label{tbl:notation}
  \vgap
  \begin{tabular*}{\linewidth}{@{\extracolsep{\fill}} p{15mm} | p{70mm}}
    \toprule
    Notation   &  Description\\
    \midrule
    $S$ &  A set of data objects\\
    $G$ & The graph index \\
    $N(x)$ & The neighbor of $x$ \\
    $o_i, q$ & A data item, query item \\
    $x_i, x_q$ & The vector of item $o_i$ and query \\
    $A_i, f_q$ & The attribute of $x_i$,  filter attributes \\
    $L_i, L_q$ & The base and query label set \\
    $[l, r]$ & the query range\\
    $N$ & The cardinality of $S$ \\
    $d$ &The dimensionality of $S$ \\
    $\delta$ & Euclidean distance \\

    \bottomrule
  \end{tabular*} \vspace{1em}
\end{table}

\begin{table}[H]
\centering
\caption{Statistics of the datasets and filter workloads.}\vgap
\label{tbl:workloads}
\small
\begin{tabular}{l|c|c|c}
\toprule
\textbf{Dataset} & \textbf{Dim} & \textbf{\#Base} & \textbf{\#Train} \\
\midrule
Tripclick-label & 768  & 1,005,976 & 49,449 \\
Youtube-label   & 128  & 1,000,000 & 1,000,000 \\
Arxiv-label     & 4096 & 1,734,264 & 1,000,000 \\
Arxiv-range     & 4096 & 1,734,264 & 1,000,000 \\
MSMARCO-range   & 1024 & 1,000,000 & 1,000,000 \\
\bottomrule
\end{tabular}
\end{table}

\begin{figure}[H]
\centering
\begin{small}

\begin{tikzpicture}
\begin{customlegend}[legend columns=2,
legend entries={$\EE$, $\EE$ w/o Filter},
legend style={at={(0.5,1.15)},anchor=north,draw=none,font=\scriptsize,column sep=0.3cm}]
\addlegendimage{line width=0.2mm,color=violate,mark=o,mark size=0.8mm}
\addlegendimage{line width=0.2mm,color=navy,mark=square,mark size=0.8mm}
\end{customlegend}
\end{tikzpicture}
\\[-\lineskip]

\begin{tikzpicture}[scale=1]
\begin{semilogxaxis}[
    width=0.7\linewidth, height=0.35\linewidth,
    xlabel=\# training queries, xlabel style={yshift=+3pt},
    ylabel=Test $R^2$, ylabel style={yshift=-4pt},
    label style={font=\scriptsize},
    tick label style={font=\scriptsize},
    xtick={1000,10000,100000,1000000},
    xticklabels={1k,10k,100k,1M},
    ymin=0.32, ymax=0.45,
    ymajorgrids=true, xmajorgrids=true,
    grid style=dashed,
]
\addplot[line width=0.2mm, color=violate, mark=o, mark size=0.6mm]
plot coordinates {
    (1000,0.370) (5000,0.411) (10000,0.419) (50000,0.431) (100000,0.434) (500000,0.431) (1000000,0.433)
};
\addplot[line width=0.2mm, color=navy, mark=square, mark size=0.6mm]
plot coordinates {
    (1000,0.332) (5000,0.362) (10000,0.372) (50000,0.389) (100000,0.392) (500000,0.389) (1000000,0.388)
};
\end{semilogxaxis}
\end{tikzpicture}

\end{small}
\vspace{-0.4cm}
\caption{Convergence of test $R^2$ vs.\ training-set size on Arxiv-contain ($k{=}10$, $f{=}81$, 1{,}000 test queries).}
\label{fig:convergence}
\end{figure}
\begin{figure}[H]

\centering
\begin{small}

\begin{tikzpicture}
\begin{customlegend}[legend columns=3,
legend entries={{Naive HNSW},{$\EE$ (10M-trained)},{$\EE$ (1M$\rightarrow$10M transfer)}},
legend style={at={(0.5,1.15)},anchor=north,draw=none,font=\scriptsize,column sep=0.3cm}]
\addlegendimage{line width=0.2mm,color=amaranth,mark=triangle,mark size=0.8mm}
\addlegendimage{line width=0.2mm,color=violate,mark=o,mark size=0.8mm}
\addlegendimage{line width=0.2mm,color=navy,mark=square,mark size=0.8mm}
\end{customlegend}
\end{tikzpicture}
\\[-\lineskip]

\begin{tikzpicture}[scale=1]
\begin{axis}[
    width=0.7\linewidth, height=0.34\linewidth,
    xlabel={NDC ($\times 10^3$)}, xlabel style={yshift=+3pt},
    ylabel=Recall@10, ylabel style={yshift=-4pt},
    label style={font=\scriptsize},
    tick label style={font=\scriptsize},
    ymajorgrids=true, xmajorgrids=true,
    grid style=dashed,
]
\addplot[line width=0.2mm, color=amaranth, mark=triangle, mark size=0.6mm]
plot coordinates {(14.258,90) (20.760,95) (30.184,97) (34.952,98) (53.666,99) (70.958,99.4)};
\addplot[line width=0.2mm, color=violate, mark=o, mark size=0.6mm]
plot coordinates {(7.483,90) (10.988,95) (14.526,97) (21.554,98) (32.072,99) (39.173,99.4)};
\addplot[line width=0.2mm, color=navy, mark=square, mark size=0.6mm]
plot coordinates {(8.142,90) (12.065,95) (15.889,97) (23.751,98) (31.479,99) (39.269,99.4)};
\end{axis}
\end{tikzpicture}

\end{small}
\vspace{-0.4cm}
\caption{NDC vs.\ Recall@10 on YFCC10M (10M vectors).}

\label{fig:yfcc-scalability}
\end{figure}
\begin{figure}[H]
    \centering
    \begin{small}

    \begin{tikzpicture}
    \begin{customlegend}[legend columns=3,
        legend entries={$\EE$ (LightGBM), Naive HNSW, $\EE$-RF},
        legend style={at={(0.5,1.18)},anchor=north,draw=none,
                      font=\scriptsize,column sep=0.3cm}]
    \addlegendimage{line width=0.2mm,color=violate,mark=o,mark size=0.8mm}
    \addlegendimage{line width=0.2mm,color=amaranth,mark=triangle,mark size=0.8mm}
    \addlegendimage{line width=0.2mm,color=navy,mark=square,mark size=0.8mm}
    \end{customlegend}
    \end{tikzpicture}
    \\[-\lineskip]

    \subfloat[Query Latency (ms)]{\vspace{-2mm}
    \begin{tikzpicture}[scale=1]
    \begin{axis}[
        height=0.15\textwidth,
        width=0.48\columnwidth,
        xlabel=Query Latency (ms), xlabel style={yshift=+3pt},
        ylabel=Recall@10, ylabel style={yshift=-4pt},
        label style={font=\scriptsize},
        tick label style={font=\scriptsize},
        ymajorgrids=true,
        xmajorgrids=true,
        grid style=dashed,
    ]
    \addplot[line width=0.2mm,color=violate,mark=o,mark size=0.6mm]
    plot coordinates {
        (1.7420, 82.78)
        (2.8690, 87.83)
        (3.2610, 90.32)
        (4.2220, 91.92)
        (5.2710, 93.09)
        (7.2320, 93.74)
        (10.2800, 94.93)
        (15.6350, 95.51)
        (21.0880, 96.05)
    };
    \addplot[line width=0.2mm,color=amaranth,mark=triangle,mark size=0.6mm]
    plot coordinates {
        (11.1270, 88.19)
        (20.6160, 92.18)
        (26.5960, 93.68)
        (32.8280, 94.55)
        (37.7810, 95.33)
        (43.3050, 95.54)
        (48.4180, 95.64)
        (52.4120, 95.78)
    };
    \addplot[line width=0.15mm,color=navy,mark=square,mark size=0.5mm]
    plot coordinates {
        (3.2720, 86.58)
        (5.6620, 90.63)
        (8.3890, 92.50)
        (13.4080, 94.51)
        (18.4300, 95.12)
        (26.9320, 95.64)
        (41.3700, 95.98)
        (59.4850, 96.27)
    };
    \end{axis}
    \end{tikzpicture}}\hspace{1mm}%
    %
    \subfloat[Avg. Distance Computation]{\vspace{-2mm}
    \begin{tikzpicture}[scale=1]
    \begin{axis}[
        height=0.15\textwidth,
        width=0.48\columnwidth,
        xlabel=Average Distance Computation, xlabel style={yshift=+3pt},
        ylabel=Recall@10, ylabel style={yshift=-4pt},
        label style={font=\scriptsize},
        tick label style={font=\scriptsize},
        ymajorgrids=true,
        xmajorgrids=true,
        grid style=dashed,
        scaled x ticks=real:10000,
        xtick scale label code/.code={$\cdot 10^4$}
    ]
    \addplot[line width=0.2mm,color=violate,mark=o,mark size=0.6mm]
    plot coordinates {
        (2006.5300, 82.78)
        (3639.5700, 87.83)
        (5246.4400, 90.32)
        (6856.9400, 91.92)
        (8537.4600, 93.09)
        (11698.4000, 93.74)
        (16519.9000, 94.93)
        (24748.3000, 95.51)
        (32931.3000, 96.05)
    };
    \addplot[line width=0.2mm,color=amaranth,mark=triangle,mark size=0.6mm]
    plot coordinates {
        (18112.6000, 88.19)
        (31519.2000, 92.18)
        (41457.1000, 93.68)
        (50577.4000, 94.55)
        (58932.9000, 95.33)
        (66932.5000, 95.54)
        (74211.8000, 95.64)
        (80002.5000, 95.78)
    };
    \addplot[line width=0.15mm,color=navy,mark=square,mark size=0.5mm]
    plot coordinates {
        (4568.9200, 86.58)
        (8758.3600, 90.63)
        (12948.5000, 92.50)
        (21331.1000, 94.51)
        (29719.8000, 95.12)
        (42266.4000, 95.64)
        (63225.3000, 95.98)
        (84198.9000, 96.27)
    };
    \end{axis}
    \end{tikzpicture}}%

    \vspace{-0.3cm}
    \caption{Alternative-predictor robustness on Tripclick-label-contain.}
    \label{fig:alt-predictor-tripclick-contain}
    \end{small}
\end{figure}
\begin{figure}[H]
    \centering
    \begin{small}

    \begin{tikzpicture}
    \begin{customlegend}[legend columns=2,
    legend entries={\EE-HNSW, DARTH-HNSW},
    legend style={at={(0.5,1.15)},anchor=north,draw=none,font=\scriptsize,column sep=0.3cm}]
    \addlegendimage{line width=0.2mm,color=violate,mark=o,mark size=0.8mm}
    \addlegendimage{line width=0.2mm,color=amaranth,mark=triangle,mark size=0.8mm}
    \end{customlegend}
    \end{tikzpicture}
    \\[-\lineskip]

    \subfloat[Arxiv-label-contain]{\vspace{-2mm}
    \makebox[0.19\textwidth][c]{%
    \begin{tikzpicture}[scale=1, trim axis left, trim axis right]
    \begin{axis}[
        height=0.07\textwidth,
        width=0.15\textwidth,
        scale only axis,
        xmode=log,
        log basis x=10,
        xlabel=Average Distance Computation, xlabel style={yshift=+3pt},
        ylabel=Recall@10, ylabel style={yshift=-4pt},
        ymin=80, ymax=100,
        label style={font=\scriptsize},
        tick label style={font=\scriptsize},
        ymajorgrids=true,
        xmajorgrids=true,
        grid style=dashed,
    ]
    \addplot[line width=0.2mm,color=violate,mark=o,mark size=0.6mm]
    plot coordinates {
        (315.2780, 76.6200)
        (429.7080, 88.1200)
        (658.5820, 95.3800)
        (887.4690, 97.6600)
        (1345.2200, 99.2100)
        (1802.9700, 99.6900)
        (2489.6100, 99.8700)
    };
    \addplot[line width=0.2mm,color=amaranth,mark=triangle,mark size=0.6mm]
    plot coordinates {
        (339.1260, 78.4500)
        (473.2530, 89.7000)
        (735.4140, 96.2100)
        (58866.4000, 100.0000)
        (177690.0000, 100.0000)
    };
    \end{axis}
    \end{tikzpicture}}}%
    \hspace{1mm}%
    \subfloat[Tripclick-label-contain]{\vspace{-2mm}
    \makebox[0.19\textwidth][c]{%
    \begin{tikzpicture}[scale=1, trim axis left, trim axis right]
    \begin{axis}[
        height=0.07\textwidth,
        width=0.15\textwidth,
        scale only axis,
        scaled x ticks=real:10000,
        xtick scale label code/.code={$\cdot 10^4$},
        xlabel=Average Distance Computation, xlabel style={yshift=+3pt},
        ylabel=Recall@10, ylabel style={yshift=-4pt},
        ymin=80, ymax=100,
        label style={font=\scriptsize},
        tick label style={font=\scriptsize},
        ymajorgrids=true,
        xmajorgrids=true,
        grid style=dashed,
    ]
    \addplot[line width=0.2mm,color=violate,mark=o,mark size=0.6mm]
    plot coordinates {
        (2006.5300, 82.7800)
        (3639.5700, 87.8300)
        (5246.4400, 90.3200)
        (6856.9400, 91.9200)
        (8537.4600, 93.0900)
        (11698.4000, 93.7400)
        (16519.9000, 94.9300)
        (24748.3000, 95.5100)
        (32931.3000, 96.0500)
    };
    \addplot[line width=0.2mm,color=amaranth,mark=triangle,mark size=0.6mm]
    plot coordinates {
        (3142.1000, 81.4600)
        (4063.0200, 82.6300)
        (4947.5500, 84.4400)
        (6898.5600, 87.7400)
        (15367.2000, 94.4400)
        (69464.7000, 96.3300)
    };
    \end{axis}
    \end{tikzpicture}}}%
    \hspace{1mm}%

    \end{small}
    \vspace{-0.35cm}
    \caption{Average NDCs for \EE-HNSW vs.\ our reimplemented DARTH-HNSW baseline.}
    \label{fig:appendix-darth-vs-e2e}
\end{figure}

\subsection*{A.2 More About Experiment Settings}\label{sec:extra-exp-set}

\stitle{Filter Construction.}
In filtered AKNN search, each vector is associated with either a categorical (label) or numerical (range) attribute, and each query is issued with an attribute predicate (i.e., a filter).
Below, we describe how we construct attributes and query filters for our workloads.

\noindent
$\bullet$
(1) \textbf{Label Attributes} (Tripclick, Arxiv, and YoutubeAudio).
For \textit{Tripclick}, we use the clinical-area labels processed by ACORN~\cite{Acorn-SIGMOD-2024}.
For \textit{Arxiv}~\cite{Arxiv-Dataset-2025-Filter-Search} and \textit{YoutubeAudio}~\cite{Youtube8M-Arxiv-2016}, we directly use the native multi-label tags (e.g., CS categories).
To study the effect of predicate strictness without confounding distribution shifts, we use the same attribute space for both containment and equality workloads.
Specifically, on \textit{Tripclick} and \textit{Arxiv}, we construct label-equality queries by enforcing exact set matching over the same multi-label distributions used for containment.
This provides a controlled stress test, where the workload becomes substantially sparser solely due to the change in predicate semantics.

\noindent
$\bullet$
(2) \textbf{Range Attributes} (Arxiv and MSMARCO).
For \textit{Arxiv}, we use native numerical metadata (e.g., update date).
For \textit{MSMARCO}, which does not provide range attributes, we follow~\cite{DIGRA-SIGMOD-2025-sibo-menxu-cuhk} and assign a synthetic integer attribute $a_i \sim \mathrm{Unif}([1,10^4])$ to each object.
Query ranges $[l,r]$ are generated to precisely control the selectivity spectrum $\{1\%, 5\%, 10\%, 20\%\}$, enabling a controlled evaluation of the cost estimator under varying filter tightness.

\stitle{Filter Workloads.}
We construct the training and evaluation workloads following Table~\ref{tbl:workloads}.

\subsection*{A.3 Adaptation to Other AKNN Methods}
\label{sec:appendix_acorn}

Our method, $\EE$, is designed to be \emph{index-agnostic}. 
Although our main evaluation focuses on standard $\POST$-filtering indices (e.g., HNSW), $\EE$ can also be applied to filter-aware indices such as ACORN~\cite{Acorn-SIGMOD-2024}. 
Yet, extending $\EE$ to such indices requires adapting the feature design to their distinct traversal mechanisms. 
We use ACORN as a representative example to illustrate this process.

\stitle{Features under $\PRE$.}
ACORN follows a $\PRE$-filtering strategy: it navigates the predicate-induced subgraph and maintains connectivity by expanding 2-hop neighbors at runtime (controlled by the parameter $\gamma$). 
A key characteristic of this traversal is that the candidate queue $\mathcal{Q}$ contains \emph{only} filter-valid nodes. 
As a result, the queue validity ratio $\rho_{queue}$ is always 1. 
Therefore, directly reusing the $\POST$-oriented features would misleadingly predict low cost for all queries.

\noindent\textbf{Feature Adaptation.}
To address this issue, we observe that for $\PRE$-filtering indices like ACORN, the dominant cost shifts from \emph{queue refinement} to \emph{neighbor inspection}. 
In particular, a \emph{hard} query in ACORN does not necessarily yield a sparse queue; instead, it incurs a high rejection rate when expanding neighbors to preserve a valid candidate set.

Accordingly, when integrating $\EE$ with ACORN, our feature extraction (i.e., $z_q$ in Algorithm~\ref{algo:search-algorithm-integration}) emphasizes \emph{Visited Correlation} rather than Queue Correlation. 
We introduce the \textbf{Visited Filter Ratio} $\rho_{visited}$:
\begin{equation}
    \rho_{visited} = \frac{N_{valid\_visited}}{N_{total\_visited}}
\end{equation}
where $N_{total\_visited}$ counts all inspected nodes during expansion, including both 1-hop and 2-hop neighbors.
A low $\rho_{visited}$ indicates that the search must traverse many invalid neighbors to sustain a valid queue, implying a large remaining budget. 
As shown in our evaluation, $\EE$ captures this implicit cost signal, enabling accurate early termination even when the explicit queue validity remains high.

\subsection*{A.4 Additional Experimental Results}
\label{sec:appendix_exp}

\stitle{Exp-4: Generalization to Other AKNN Methods.}\label{exp:general-to-sota}
In Figure \ref{fig:recall-latency-acorn-gamma}, we show that \EE generalizes well to ACORN's $\PRE$ strategy, and can accurately capture the cost of the subsequent expansion stage via the \textit{Visited Filter Ratio} ($\rho_{visited}$).
As a result, \EE improves the search efficiency of ACORN-$\gamma$ by up to $1.5\times$, highlighting the robustness and portability of our adaptive termination framework.

\stitle{Exp-5: Convergence on training samples.}  We change the training-set size from 1k to 1M queries on Arxiv-contain ($K{=}10$, $f{=}81$) and report test $R^2$ on a disjoint 1k-query set (Figure~\ref{fig:convergence}). Test $R^2$ rises steeply up to $\sim$50k training queries and then plateaus. The gap between \EE and the \EE w/o Filter ablation stays at $0.04$--$0.05$ across the different training set sizes, showing that our filter-aware features consistently improve the predictor accuracy.

\stitle{Exp-6: Scalability to YFCC 10M.} To evaluate the scalability of \EE, we evaluate $\EE$ on YFCC10M---the largest real-attribute dataset we could identify. Larger benchmarks typically require synthesizing attributes, which breaks the vector--attribute correlation that motivates filter-aware features (\S~\ref{sec:problem-analysis}). As shown in Figure ~\ref{fig:yfcc-scalability}, The model trained on YFCC1M transfers to YFCC10M with only $\sim$10\% NDC overhead, with $50$k training queries and $<$5\,s training even at this scale. 

\stitle{Exp-7: Alternative Predictor.}  To verify that the gain is not specific to LightGBM, we replace it with a Random Forest cost predictor on TripClick-contain in Figure \ref{fig:alt-predictor-tripclick-contain}, showing that the filter-aware features carry the gain across predictor families.

 \stitle{Exp-8: Comparison with DARTH.}\label{exp:darth-comparison}
  To enable a direct comparison with state-of-the-art learned termination, we reimplement DARTH~\cite{DARTH-SIGMOD-2026} on hnswlib (the same
  backbone as \EE), since its original Faiss implementation does not support attribute-aware post-filtering.
  Figure~\ref{fig:appendix-darth-vs-e2e} reports NDC vs.\ recall@10 on the two label-containment datasets and confirms that \EE's feature-aware feature provides a more reliable termination signal than distance-only features.


\end{document}